\documentstyle[aps,preprint,eqsecnum]{revtex}

\def\tvij{\tilde{v}_{ij}}
\def\tv{\tilde{v}}
\def\dvij{\delta v ({\bf P}_{ij})}
\def\dv{\delta v }
\def\beq{\begin{equation}}
\def\eeq{\end{equation}}
\def\ba{\begin{eqnarray}}
\def\ea{\end{eqnarray}}
\def\vij{v_{ij}}
\def\Vijk{V_{ijk}}
\def\vet{v_{18}}
\def\v8p{v_8^\prime}
\def\vpi{v^{\pi}}
\def\Vtpip{V^{2\pi,PW}}
\def\Vtpis{V^{2\pi,SW}}
\def\Vthpi{V^{3\pi, \Delta R}}
\def\Vthpia{V^{3\pi, \Delta R}_1}
\def\Vthpib{V^{3\pi, \Delta R}_2}
\def\Otpip{O^{2\pi,PW}_{ijk}}
\def\Otpis{O^{2\pi,SW}_{ijk}}
\def\Othpi{O^{3\pi, \Delta R}_{ijk}}
\def\Othpia{O^{3\pi, \Delta R}_{1,ijk}}
\def\Othpib{O^{3\pi, \Delta R}_{2,ijk}}
\def\OR{O^R_{ijk}}
\def\Atpip{A_{2\pi}^{PW}}
\def\Atpis{A_{2\pi}^{SW}}
\def\Athpi{A_{3\pi}^{\Delta R}}
\def\Athpib{A_{2,3\pi}^{\Delta R}}
\def\AR{A_R}
\def\Vtpi{V^{2\pi}}
\def\vr{v^R}
\def\VR{V^R}
\def\rij{r_{ij}}
\def\opij{O^p_{ij}}
\def\vpij{v_p(r_{ij})}
\newcommand{\avet}{\mbox{AV18~}}
\newcommand{\avep}{\mbox{AV8$^\prime$~}}
\newcommand{\boldsigma}{\mbox{\boldmath$\sigma$}}
\newcommand{\boldtau}{\mbox{\boldmath$\tau$}}

\begin{document}

\draft
 
\tighten
{\tighten

\title{Realistic models of pion-exchange three-nucleon interactions}

\author
{ Steven C. Pieper$^{1,}$\cite{scp}, 
 V. R. Pandharipande$^{2,}$\cite{vrp}, 
  R.~B.~Wiringa$^{1,}$\cite{rbw}, 
  J. Carlson$^{3,}$\cite{jc} } 

\address
{$^1$Physics Division, Argonne National Laboratory, Argonne, Illinois 60439 \\
 $^2$Department of Physics, University of Illinois, Urbana, Illinois 61801 \\
 $^3$Theoretical Division, Los Alamos National Laboratory, Los Alamos,
         New Mexico 87545 \\}

Revised - \date{\today}

\maketitle

\begin{abstract}

We present realistic models of pion-exchange three-nucleon interactions 
obtained by fitting the energies of all the 17 bound or narrow 
states of $3 \leq A \leq 8$ nucleons,  
calculated with less than 2\% error using the Green's function Monte 
Carlo method.  The models contain two-pion-exchange terms due to 
$\pi N$ scattering in S- and P-waves, three-pion-exchange terms 
due to ring diagrams with one $\Delta$ in the intermediate states, 
and a phenomenological repulsive term to take into account relativistic 
effects, the suppression of the two-pion-exchange two-nucleon interaction by 
the third nucleon, and other effects.  The models have five parameters,
consisting of the strength of the four interactions and the short-range 
cutoff. The 17 fitted energies are insufficient to determine all of 
them uniquely. 
We consider five models, each having three adjustable parameters 
and assumed values for the other two.  They reproduce the 
observed energies with an rms error $<$ 1\% when used together with 
the Argonne $v_{18}$ two-nucleon interaction.  In one of the models 
the $\pi N$ S-wave scattering interaction is set to zero; 
in all others it is assumed to have the strength suggested by chiral 
effective field theory.  One of the models also assumes that the 
$\pi N$ P-wave scattering interaction has the strength suggested by 
effective field theories, and the cutoff is adjusted to fit the data. 
In all other models the cutoff is taken to be the same as in the 
$v_{18}$ interaction.  The effect of relativistic boost correction to 
the two-nucleon interaction on the strength of the repulsive three-nucleon 
interaction is estimated. Many calculated properties 
of $A \leq 8$ nuclei, including radii, magnetic dipole and electric 
quadrupole moments, isobaric analog energy differences, etc., are tabulated.
Results obtained with only Argonne $v_8^\prime$ and $v_{18}$ interactions 
are also reported.  In addition, we present results for 7- and 8-body 
neutron drops in external potential wells. 

\end{abstract}

\pacs{PACS numbers: 21.10.-k, 21.45.+v, 21.30.+y, 13.75.Cs, 12.40.Qq, }

}
 
\narrowtext

\section{Introduction}

One of the primary goals of nuclear physics is to understand the stability,
structure, and reactions of nuclei as a consequence of the interactions
between individual nucleons.
However, these interactions are not known from first principles; they are 
modeled with parameters to be determined from data. 
Significant advances have been made during the last decade in the 
{\it ab initio} calculation of nuclear properties starting from these realistic 
models of the nuclear force, which allow us to test the predictions of such 
models with unprecedented accuracy, and to refine them.
With our collaborators, we have carried out a series of many-body
calculations of light nuclei~\cite{PPCPW97,WPCP00} and nuclear and neutron
star matter~\cite{APR98} using a Hamiltonian that contains both two- and
three-nucleon potentials.
The light nuclei calculations use the Green's function Monte Carlo (GFMC) 
method and have been demonstrated to give nuclear binding energies for up to 
eight-body nuclei with a precision of better than 2\%.
The matter calculations are less accurate but provide important constraints 
on the Hamiltonian.
These calculations have used the Argonne $\vet$ (AV18) model~\cite{WSS95} 
of the two-nucleon interaction, $\vij$, and the Urbana~IX (UIX) 
model~\cite{PPCW95} of the three-nucleon interaction, $\Vijk$.

The results for light nuclei are summarized in Fig.~\ref{fig:oldH}, where
we compare the calculated and experimental binding energies for all the ground 
or narrow, low-lying, excited states of nuclei with up to eight nucleons 
(neglecting isobaric analog states).
In addition to the predictions of the AV18/UIX Hamiltonian, we show the
results (most newly calculated for the present paper) of using just the 
two-body AV18 interaction by itself.
We see that AV18 alone predicts some key features of nuclear structure 
correctly, such as the proper ordering of excited states and the rapid 
saturation of the binding above $^4$He.
However, with the exception of $^2$H, it underbinds all nuclei, and this
failure grows rapidly with increasing $A$.
With just the two-nucleon force acting, the Borromean nuclei $^6$He and $^8$He 
are not stable and the lithium nuclei are only marginally so.

The addition of the UIX model of $\Vijk$ fixes the binding energy of $^3$H
and $^4$He and significantly improves the binding of the p-shell nuclei.
However, AV18/UIX still underbinds as $A$ increases, and also as $N-Z$ 
increases.
In particular, $^8$He is more underbound than $^8$Be, indicating a 
problem with the isospin dependence of this interaction model. 
The relative stability of the lithium nuclei is improved, but 
the Borromean helium nuclei are still unbound.
Additional calculations of wider, higher-lying, excited states not shown in
Fig.~\ref{fig:oldH} indicate another problem with the AV18/UIX
model: the underprediction of spin-orbit splittings among spin-orbit
partners such as the $\case{3}{2}^-$ and $\case{1}{2}^-$ states in $^5$He.

In this paper we investigate new models of $\Vijk$ that largely correct these
failings and give a very good description of the spectrum of light nuclei.
Studies of nuclear and neutron star matter with these new models will be
reported in a separate paper.

The theory of strong interactions has not yet progressed enough to permit 
a first-principles determination of the two- and three-nucleon interactions 
with the accuracy required to calculate nuclear binding energies. 
The interactions must be determined phenomenologically.
Modern, realistic models of $\vij$ are obtained by fitting the 
$\sim$4300 data below 350 MeV in the Nijmegen 
$N\!N$-scattering data base~\cite{SKRD93} with a $\chi^2 \sim 1$ per degree 
of freedom.  The Nijmegen data base is said to be complete, i.e.,
the included data determine all the relevant phase shifts and mixing 
parameters.  Thus $\vij$ fitted to it are well determined and generally
give very similar predictions of the properties of three- and four-body
nuclei, as will be discussed below.

In contrast it is much more difficult to construct realistic models of
$\Vijk$ by simply fitting three-nucleon scattering data, which is dominated
by the pairwise forces.   
The number of operators that can contribute to $\Vijk$ is very large,
and until recently, the number of observables that could both be observed
and accurately calculated was small.
Recent advances in three-nucleon scattering calculations, based on correlated 
hyperspherical harmonic~\cite{Ketal01} and Faddeev~\cite{Wetal01} methods,
and in high-precision $Nd$ scattering experiments, hold significant promise
for testing models of $\Vijk$ in this regime.
However, the binding energies and 
excitation spectra of light nuclei also contain a great deal
of information, and are in fact the only current means to investigate
$T=\case{3}{2}$ forces.

An additional concern is that the $\Vijk$ obtained by fitting nuclear data may 
depend strongly on the model of $\vij$ used in the Hamiltonian.
The $\Vijk$ will naturally depend upon the chosen $\vij$ to some extent.
For example, two equivalent models of $\vij$, related by a unitary 
transformation, will have different but related $\Vijk$ associated with them 
\cite{coonf}.
However, combinations of $\vij$ and $\Vijk$ related by unitary 
transformations will naturally predict the same observables. 

Models of $\Vijk$ based on the elimination of field variables date back to the work
of Primakoff and Holstein~\cite{PH39}.
The first modern meson-exchange 
model for nuclear $\Vijk$ was proposed by Fujita and Miyazawa (FM)~\cite{FM57};
it contained only the two-pion-exchange three-nucleon interaction $\Vtpip$ 
due to scattering of the pion being exchanged between two nucleons by a third 
nucleon via the P-wave $\Delta$-resonance. 
This interaction is attractive in nuclei and nuclear matter.
Later theoretical models, such 
as Tucson-Melbourne (TM)~\cite{TM} and Brazil~\cite{brazil} 
included the $\Vtpis$ due to $\pi N$ S-wave scattering and $\Vtpip$ 
from all P-wave scattering.  In the recent Texas model these two-pion-exchange 
contributions to $\Vijk$ have been predicted using chiral symmetry
\cite{fhk99}.  The FM and later models have 
similar forms for $\Vtpip$, but the predicted strength of the long-range 
part of $\Vtpip$ in the later models is almost twice that in FM.  

The main failures of a nuclear Hamiltonian containing only two-nucleon 
interactions include the underbinding of light nuclei, as discussed above, 
and an overestimate of the equilibrium
density of nuclear matter.  An attractive $\Vtpi$ addresses the
first failure while making the second worse~\cite{CPW83}.
The Urbana models of $\Vijk$ contain only two terms, the $\Vtpip$ and 
a phenomenological, repulsive $\VR$.
The strengths of the two interactions in the most recent Urbana model, UIX,
were obtained by reproducing the energy of $^3$H via a
GFMC calculation and the density of nuclear
matter by approximate variational calculations~\cite{PPCW95,APR98}.  
The repulsive term $\VR$ in $\Vijk$ is essential
to prevent nuclear matter from being too 
dense and overbound.  

The expectation value of the $N\!N\!N$ potential is much smaller than 
that of the $N\!N$ potential in nuclei. For example, the ratio
of contributions of the UIX and AV18 potentials in $A\leq8$ nuclei
is $< 0.1$~\cite{PPCPW97,WPCP00}.
However, the $\Vijk$ gives a relatively much larger contribution
to nuclear binding energies due to the significant cancellation between the
positive kinetic energy and the negative $N\!N$ potential.  
It is this feature that allows us extract information on $\Vijk$ by
studying the spectrum of light nuclei.

In the present work we fit the energies of 17 states of up to eight 
nucleons, calculated by the GFMC method with $<~2\%$ error, to construct 
more realistic models of $\Vijk$.  These models are for use with the AV18 
$\vij$. 
In addition to the already mentioned $\Vtpip$, $\Vtpis$, and $\VR$ terms, 
they contain three-pion-exchange rings with 
$\Delta$ intermediate states, $\Vthpi$.  All the terms 
are static; their spin-isospin and spatial dependence is taken from 
theoretical models, and their strengths are varied to fit the
observed energies. 

The new models are refered to as ``Illinois'' models; five versions,
Illinois-1 to -5 (designated IL1 to IL5) are presented in this paper.
The Hamiltonians using AV18 and these $\Vijk$ are refered to as AV18/IL1, etc.
For each model, two to three of the available five parameters were adjusted
to fit the binding energies of the 17 states assuming plausible values for 
the other parameters.
The IL1 and IL2 models have short-range cutoffs taken from \avet, 
while IL3 uses the strength predicted by chiral perturbation theory~\cite{fhk99}
for $V^{2 \pi}_{ijk}$, and adjusts the cutoff 
to fit the energies.  IL4 and IL5 are further variations of IL2.  
The qualities of the fits
are good, and the extracted strength parameters 
have plausible values.  This  suggests that
strengths of additional terms 
in $\Vijk$ cannot be determined from the data included in the present work.
It is also 
possible that additional terms in $\Vijk$ are weaker than $\Vtpis$ 
and $\Vthpi$, which in turn are weaker than the $\Vtpip$ and $\VR$ 
considered in the older Urbana models.  

Several relativistic effects are contained in the two- and three-nucleon 
potentials fitted to experimental data.  However, 
the boost correction, $\dvij$, to the two-body interaction is omitted in 
nonrelativistic Hamiltonians containing $v_{ij}$ fitted to the scattering data 
in the two-nucleon center of mass frame.  
This many-body effect arises from the
motion of the center of mass of the $ij$ pair of nucleons in the presence
of the other nucleons.
Initially the boost interaction $\dvij$ is neglected in the GFMC calculations.
It is subsequently treated as a first-order perturbation.  The contribution
of $\dvij$ to the binding energy of light nuclei is nearly proportional to that
of $\VR$.  The final value of the strength of $\VR$ can be adjusted
to reproduce the observed energies when the perturbatively computed
$\dvij$ contribution is included.  

A brief review of the $N\!N$ interaction, including relativistic corrections,
is given in Sec. II.  The new Illinois models of $\Vijk$ are presented
in Sec.~III.
The GFMC calculations of light nuclei 
are briefly described in Sec. IV.  The nuclear energies calculated with 
\avet and its approximation AV8$^{\prime}$, as well as those including 
the new Illinois three-nucleon potentials are reported in Sec. V. 
A number of results obtained with the new Illinois models for the light 
nuclei, including proton and neutron distribution radii, magnetic and
quadrupole moments, and isobaric analog energy differences, are also given 
in Sec. V.  
In addition, we report results obtained for drops of 7 and 8 neutrons in an 
external potential well to provide constraints for energy-density functionals 
of neutron-rich nuclei~\cite{PSCPPR96}. 
Our conclusions are given in Sec. VI. 

\section{The Two-nucleon Interaction}

We use the Hamiltonian:
\begin{equation}
H = \sum_i - \frac{\hbar^2}{2m_i} \nabla_i^2 + \sum_{i < j} \vij 
  + \sum_{i<j<k} \Vijk \ ,
\label{eq:hnr}
\end{equation}
containing kinetic, two- and three-nucleon interaction energies. 
The mass difference between the proton and the
neutron is taken into account by letting $m_i$ be the mass of proton or 
neutron according to the isospin of nucleon $i$, and both strong and electromagnetic
isovector and isotensor terms
are included in the $\vij$.  

The Argonne $\vet$ two-nucleon potential~\cite{WSS95} contains $v^\pi$,
the one-pion-exchange potential with a short-range cutoff, $v^R$
representing all other strong interaction terms, and $v^\gamma$, a
very complete treatment of the electromagnetic interaction:
\begin{equation}
\vij = v^\pi_{ij} + v^R_{ij} + v^\gamma_{ij} \ .
\end{equation} 
It can be expressed as a sum: 
\begin{equation}
\vij = \sum_p \ \vpij \opij \ ,
\label{eq:vijsum}
\end{equation} 
in which $\opij$ are operators, and $\vpij$ depend only on the interparticle 
distance $\rij$.  The first six operators are the only possible 
isospin-conserving static ones, i.e., operators independent of the 
nucleon velocities:
\begin{equation}
O^{p=1,6}_{ij} = \left( 1,\ \boldsigma_i \cdot \boldsigma_j ,\ S_{ij}
\right) \otimes \left( 1,\ \boldtau_i \cdot \boldtau_j \right)\ ,
\label{eq:stos}
\end{equation}
where $S_{ij}$ is the two-nucleon tensor operator.  There are only two 
isospin-conserving spin-orbit terms linear in the velocities, with operators:
\begin{equation}
O^{p=7,8}_{ij} = {\bf L} \cdot {\bf S} 
\otimes \left( 1,\ \boldtau_i \cdot \boldtau_j \right)\ ,
\label{eq:soos} 
\end{equation} 
where ${\bf L}$ and ${\bf S}$ are the relative angular momentum and the 
total spin respectively.   
The above eight terms are unique and able to describe most of the features 
of the $N\!N$ interaction.  The long-range parts of $v_4(\rij)$ and 
$v_6(\rij)$, associated with the $\boldsigma_i \cdot \boldsigma_j \ 
\boldtau_i \cdot \boldtau_j $ and $S_{ij} \ \boldtau_i \cdot \boldtau_j $
operators respectively, are given by the one-pion-exchange potential $\vpi$.
In addition there are phenomenological parameterizations of the short- and 
intermediate-range parts of the $\vpij$.  

It is necessary to add several smaller terms to the above eight in order 
to fit the scattering data with a $\chi^2 \sim 1$. 
These include terms dependent quadratically on the velocity, and static 
and spin-orbit terms breaking the isospin symmetry.  In \avet
the quadratic operators are chosen as:
\begin{equation}
O^{p=9,14}_{ij} = \left( L^2,\ L^2 \boldsigma_i \cdot \boldsigma_j ,\
({\bf L} \cdot {\bf S})^2 \right)
\otimes \left( 1,\ \boldtau_i \cdot \boldtau_j \right)\ ;
\label{eq:qdos}
\end{equation}
however, in the Paris~\cite{paris} and Nijmegen~\cite{SKTS94} models 
$\nabla^2$ is used in place of the $L^2$.  

In addition to the isospin-breaking terms in $v^\gamma$, strong-interaction
isospin-breaking terms are necessary to reproduce the data.
The $O^{p=15-17}_{ij}$ are 
isotensors with central, $\boldsigma_i \cdot \boldsigma_j$, and tensor operators, 
and the long-range part of $v_{p=15-17}(\rij)$ is determined from  
the difference of the masses of charged and neutral pions.  The isovector 
term associated with $O^{p=18}_{ij}$ is necessary to explain the difference 
in the $T=1,\ S=0 \ pp$ and $nn$ scattering lengths~\cite{WSS95}. 
The interactions associated with the 18 operators listed above contain 
all the strong and parts of the electromagnetic interaction.  In addition, 
there are four more operators that appear only in the $v^\gamma$.
The number of parameters contained in the \avet model of $\vij$ is 
$\sim$ 40, and all of them are fairly well determined by the $\sim 4300$  
data in the Nijmegen data base.

It is well known that two-nucleon scattering data up to 350 MeV cannot 
determine the potential $\vij$ uniquely.  In addition
to \avet, there are four other modern models: 
Reid-93, Nijmegen-I and II~\cite{SKTS94},
and CD-Bonn~\cite{MSS96}, all of which fit the Nijmegen data base.
The five models are different from each other in detail.  The 
Reid-93, Nijmegen-II and \avet models assume that the 
interaction in each $LSJ$ partial wave can be represented by a local 
potential in that partial wave; in addition the operator structure of the
\avet model given above relates the potentials in all partial waves.
In states with total spin  $S=1$
the local potential in $LSJ$-$L^{\prime}SJ$ coupled waves is 
expressed as a sum of central, tensor and spin-orbit 
components.  On the other hand, the Nijmegen-I and CD-Bonn models 
include non-local interactions based on boson-exchange phenomenology.  

All the models of $\vij$ contain one-pion-exchange potentials, $\vpi$,
as the long-range part, and phenomenological shorter-range parts.
Fortunately the $\vpi$ gives the largest contribution to nuclear potential 
energies, and thus the model dependence of the phenomenological parts 
has a limited scope.  However, the $\vpi$ itself is not uniquely predicted 
by theory. That in the CD-Bonn model is derived assuming 
pseudoscalar pion-nucleon coupling, and is nonlocal, while that in the 
other models is essentially local.  The deuteron and $^1S_0$-scattering 
wave functions predicted by the five models are compared in Ref.~\cite
{ppcap}.  All the local models predict essentially the same wave functions, 
however, the two nonlocal models give different wave functions. 
The main difference is in the $D$-state wave function of the deuteron;
that predicted by CD-Bonn is smaller at $r < 2$ fm, while that of
Nijmegen-I is close to predictions of local models at all values of $r$.
The $S$-state wave functions of the boson-exchange models, CD-Bonn and
Nijmegen-I, are larger than those of the local models at $r < 1 $ fm.

The deuteron elastic-scattering form factors are sensitive to the 
wave function.  The $A(q^2)$ structure function has been accurately measured, 
most recently at Jefferson Lab~\cite{ahalla,ahallb}, and results of 
the recent measurements of the tensor analyzing power, $T_{20}(q^2)$, at
Jefferson Lab~\cite{t20pc1} and NIKHEF~\cite{t20pc2} are also available.  
These indicate that the deuteron wave functions calculated from the local 
potentials are very realistic.  They correctly predict the observed 
data with plausible pair currents~\cite{CS98}.  In addition, it has been 
shown recently~\cite{jundp} that nonrelativistic calculations using local
$\vpi$ give deuteron wave functions close to those predicted with nonlocal 
$\vpi$ obtained with the pseudovector pion-nucleon coupling, favored 
by chiral perturbation theories, and relativistic kinetic
energy.  The corrections of order $p^2/m^2$, to the amplitudes of states
with large momentum $p$, coming from relativistic nonlocalities of $\vpi$ 
and relativistic kinetic energy cancel in this case. 

The main assumption we make here is that local models provide an accurate
representation of $\vij$.  It is supported by the observed deuteron form
factors mentioned above, and is valid for the $\vpi$ as shown in Ref.~\cite{jundp}.
The local models also predict essentially the same value (7.63 MeV for 
Reid-93 and 7.62 MeV for \avet and Nijmegen-II) of the binding
energy of the triton in non-relativistic calculations with no three-nucleon
potential~\cite{FPSS93}.
The difference between these values
and the observed binding energy of 8.48 MeV is one of the indications for 
the presence of $\Vijk$ in the Hamiltonian (\ref{eq:hnr}).  The triton 
energies obtained with the Nijmegen-I and CD-Bonn models, without $\dvij$, 
are 7.74 and 8.01 MeV respectively~\cite{NKG00}.  The energies
of the alpha particle predicted by  \avet and Nijmegen-II differ by only 
0.28 MeV while those predicted by Nijmegen-I and CD-Bonn models are respectively
0.7 and 2.0 MeV more bound than the \avet value~\cite{NKG00}.

Accurate calculations of nuclear matter are not yet practical.  Nevertheless 
the nuclear matter equation of state has been studied for all five modern 
potentials with the lowest-order Brueckner-Hartree-Fock method with continuous 
single-particle energies~\cite{oslo}, again without relativistic corrections.  The local
interactions, Nijmegen-II, Reid-93 and \avet, give similar results, while the
most nonlocal CD-Bonn gives the lowest energies.
The predicted values of equilibrium $E_0$ and $\rho_0$ of symmetric nuclear 
matter are $-$17.6~MeV at 0.27~fm$^{-3}$ with Nijmegen-II, 
$-$18.1 at 0.27 with \avet, $-$18.7 at 0.28 with Reid-93, 
$-$20.3 at 0.31 with Nijmegen-I and $-$22.9 at 0.37 with CD-Bonn,
while the empirical values are $-16$ MeV at 0.16 fm$^{-3}$.
The above Brueckner results for \avet are quite close to the  
$E_0 = - 18.2$ MeV and $\rho_0 = 0.3$ fm$^{-3}$ obtained with the 
variational method using chain summation methods~\cite{APR98}.  
The triton and $^4$He energies obtained with the nonlocal CD-Bonn interaction 
are closer to experiment than the predictions of local models, 
but its predicted nuclear matter properties are farther away. 

It has been stressed by Friar~\cite{friaropep} that the various representations 
of $\vpi$ are related by unitary transformations.  It should be
possible to use these transformations to find the appropriate current operators 
that will explain the deuteron form factors with wave functions predicted by 
the nonlocal models.  These transformations will also generate three-body 
forces accounting for the difference between energies obtained from 
local and nonlocal models.  Thus the deuteron form factors do not exclude 
nonlocal representations of $\vij$.  However, it seems that the simplest 
realistic models of the nuclear Hamiltonian may be obtained with local $\vij$, 
and fortunately there is much less model dependence in these.  In the 
present work we use the \avet model of $\vij$; however, the other 
local models will presumably require similar $\Vijk$.

The two-nucleon interaction $\vij$
depends both on the relative momentum ${\bf p} = ({\bf p}_i - {\bf p}_j)/2$
and the total momentum ${\bf P} = {\bf p}_i + {\bf p}_j $ of the 
interacting nucleons.  We can express it as:
\begin{equation}
\vij = \tvij + \dvij \ ,
\label{eq:tvdv}
\end{equation}
where $\delta v ({\bf P}=0)=0$.  The models discussed above give $\tvij$ in the ${\bf P}=0$,
center of momentum frame.  In many calculations the $\tvij$ is used as an
approximation to $\vij$ by neglecting the boost correction $\dvij$.  In
fact terms dependent on ${\bf p}$ included in $\tvij$ 
are of the same order as those in $\dvij$ dependent on ${\bf P}$~\cite{FPF95}.  
It is essential to include the $\dvij$ to obtain the true momentum dependence 
of the $\vij$.  For example, the electromagnetic interaction between two 
charges, as well as the analogous vector-meson-exchange interaction between 
two nucleons depends upon 
${\bf p}_1 \cdot {\bf p}_2 = \case{1}{4}{\bf P}^2 - {\bf p}^2$. 
The $\tv$ includes only the ${\bf p}^2$ term, while the ${\bf P}^2$ term is 
in $\dv$.  The $\dv$ is related to $\tv$ and its leading term
of order ${\bf P}^2$ is given by:
\begin{equation}
\delta{v}({\bf P})=-\frac{P^2}{8m^2}\tilde{v}+\frac{1}{8m^2}\,\left[\ {\bf P}
\cdot{\bf r}\;{\bf P}\cdot\mbox{\boldmath$\nabla$}, \tilde{v}\ \right]+
\frac{1}{8m^2}\,\left[\ (\mbox{\boldmath$\sigma$}_1-\mbox{\boldmath$\sigma$}_2)
\times{\bf P}\cdot\mbox{\boldmath$\nabla$}, \tilde{v}\ \right]\ .
\label{eq:friar}
\end{equation}
The validity of the above equation, obtained by Friar~\cite{F75}, in classical and
quantum relativistic mechanics and in relativistic field theory has been
shown in Ref.~\cite{FPF95}.

The effects of the $\dvij$ on the energies of $^3$H and $^4$He~\cite{rel2} and 
nuclear matter~\cite{APR98} have been studied for the AV18 model using the 
variational method. 
This boost correction gives a repulsive contribution in both cases.
It increases the triton energy by $\sim$ 0.4 MeV away from experiment, while 
the nuclear matter equilibrium $E_0$ and $\rho_0$ move to $-13.7$ MeV at
0.23 fm$^{-3}$, which is closer to the empirical density, but farther from
the empirical energy.
The VMC studies~\cite{rel2} of $\dvij$ also show that the dominant
corrections come from the first and second terms of Eq.(\ref{eq:friar})
and that only the first six operator terms (the static terms) of AV18
give substantial contributions.  Accordingly, we ignore the last term
of Eq.(\ref{eq:friar}) in this paper and evaluate the first two for only
the static parts of $\tv$.  Furthermore it was shown that the terms
arising from the derivatives acting on operators in $\tv$ were
negligible, so we do not evaluate them here.

\section{Illinois Models of $\Vijk$ }

The Illinois $\Vijk$ are expressed as:
\begin{equation}
\Vijk = \Atpip \Otpip + \Atpis \Otpis + \Athpi \Othpi + \AR \OR \ .
\label{eq:vijk}
\end{equation}
Their four terms represent the $\Vtpip,\Vtpis,\Vthpi$, and $\VR$ interactions 
with strengths $\Atpip,\Atpis,\Athpi$, and $\AR$. In the 
following subsections we give the spin-isospin and spatial operators 
associated with these interactions and the theoretical estimates of 
the strengths.  In the older Urbana models $\Atpip$ is denoted by $A_{2 \pi}$, 
$A_R$ by $U_0$ and the $\Vtpis$ and $\Vthpi$ terms are absent.

\subsection{$\Vtpip$}

The earliest model of $\Vtpip$ is due to Fujita and Miyazawa~\cite{FM57}, who
assumed that it is entirely due to the excitation of the $\Delta$-resonance
as shown in Fig.~\ref{fig:vijk}a.  Neglecting the nucleon and $\Delta$ kinetic
energies we obtain:
\begin{eqnarray}
\Atpip &=& - \frac{2}{81} \  \frac{f^2_{\pi NN}}{4 \pi} \  
\frac{f^2_{\pi N \Delta}}{4 \pi} \  \frac{m_{\pi}^2}{(m_{\Delta}-m_N)} \ ,
\label{eq:atpip}     \\
\Otpip &=& \sum_{cyc} \left( \left\{ X_{ij}, X_{jk} \right\} 
\left\{ \boldtau_i \cdot \boldtau_j , \boldtau_j \cdot \boldtau_k \right\} 
+ \case{1}{4}  \left[ X_{ij}, X_{jk} \right] 
\left[ \boldtau_i \cdot \boldtau_j , \boldtau_j \cdot \boldtau_k \right] \right) \ , 
\label{eq:otpip}   \\
X_{ij} &=& T(m_{\pi}r_{ij}) \  S_{ij} + Y(m_{\pi}r_{ij}) \  
\boldsigma_i \cdot \boldsigma_j \ , 
\label{eq:xij}    \\
Y(x) &=& \frac{e^{-x}}{x} \  \xi_Y(r) \ , 
\label{eq:yij}   \\
T(x) &=& \left( \frac{3}{x^2} + \frac{3}{x} + 1 \right) Y(x) \  \xi_T(r) \ . 
\label{eq:tij}
\end{eqnarray}
Here $\xi_Y(r)$ and $\xi_T(r)$ are short-range cutoff functions.  We note 
that the one-pion-exchange two-nucleon interaction used in AV18 is given by:
\begin{equation}
v^{\pi}_{ij} = \frac{1}{3} \  \frac{f^2_{\pi NN}}{4 \pi} \  m_{\pi} \  
 \boldtau_i \cdot \boldtau_j \ X_{ij} \ ,
\label{eq:vpi}
\end{equation}
with  cutoff functions
\begin{equation}
\xi_Y(r) = \xi_T(r) = (1- e^{-c r^2}) \ ,
\label{eq:cutoff}
\end{equation}
and $c = 2.1 $ fm$^{-2}$.  The contact, $\delta$-function part, of the 
OPEP is not included in the $\vpi$ in Urbana-Argonne models since it is 
difficult to separate it from the other short-range parts.
These functional forms are used in UIX and all the Illinois models.  

In all the Illinois models except IL3 the cutoff $c = 2.1$~fm$^{-2}$ is 
used and the $\Atpip$ is varied to fit the data, as in UIX.  This
approximation assumes that the $\pi N\!\Delta$ form factor is similar to 
the $\pi N\!N$ form factor.  In fact it is likely that the radius of the 
$\Delta$-resonance is larger than that of the nucleon, and thus the 
 $\pi N\!\Delta$  form factor is softer than the $\pi N\!N$.  In this case 
 use of the $T(x)$ and $Y(x)$ functions from $\vij$ in $\Vtpip$ would 
lead to an underestimation of $\Atpip$.  In the IL3 model we use 
a value of $\Atpip$ typical of the Tucson, Brazil, and Texas models
\cite{TM,brazil,fhk99} and vary the cutoff parameter $c$ in $\Vijk$ to fit 
the data.  

Using the observed values of $m_{\Delta}$ and $f^2_{\pi N \Delta}/4\pi 
\sim 0.3$, Eq.(\ref{eq:atpip}) predicts that $\Atpip \sim - 0.04$ MeV.
With the cutoffs from $\vij$, the $\Vtpip$ of this strength
gives a contribution of $\sim - 3$ MeV
to the energy of $^3$H.  It is much larger than the $-0.9$ to $-0.6$ MeV 
estimated by Faddeev calculations~\cite{sauer,prb} 
that include explicit $\Delta$ degrees
of freedom and $N\!N \rightleftharpoons N\!\Delta$
transition potentials~\cite{WSA84}.  A part of the difference is 
probably due to the neglect of kinetic energies of the nucleons and 
$\Delta$ in the energy denominator in Eq.(\ref{eq:atpip}). 
Neglecting the momenta of the  
nucleons before the pion emission, the energy denominator in 
Eq.(\ref{eq:atpip}) should be $m_{\Delta} - m_N + q_{\pi}^2(1/2m_{\Delta} 
+ 1/2m_N)$, where ${\bf q}_{\pi}$ is the momentum of the first pion in 
Fig.~\ref{fig:vijk}a.  The average momenta of pions exchanged in interactions
between nucleons in nuclei is $\sim 500$ MeV/c~\cite{fpw,AP97}, for which 
Eq.(\ref{eq:atpip}) underestimates the denominator by $\sim 40\%$.  It 
thus appears likely that Eq.(\ref{eq:atpip}) overestimates the strength 
of $\Vtpip$ via the $\Delta$-resonance significantly.  

The other models of $\Vijk$ start from the observed pion-nucleon 
scattering amplitude, and use current algebra and PCAC constraints, 
or chiral symmetry, to 
extrapolate to the off-mass-shell pions responsible for the $\Vtpi$.  In 
this way they include the contributions of all the $\pi N$ resonances, 
as well as that of $\pi N$ S-wave scattering to the $\Vtpi$.  The TM 
$\Vtpi$ has been cast in the form of Eq.(\ref{eq:vijk}) in Ref.~\cite{CPW83}.
It contains the term $\Vtpip$ with the operator $\Otpip$ and the strength 
$\Atpip = -0.063$ MeV.  The strength $\Atpip$  is proportional to the 
parameter $b$ in the $\pi N$ scattering amplitude, and the values of 
$b$ in various models have been tabulated by Friar et al.~\cite{fhk99}. 
The Texas model has the largest value of $b$ corresponding to 
$\Atpip = -0.09$.  These strengths are much larger than $- 0.04$ 
estimated with the simple Fujita-Miyazawa model presumably because of 
the additional contributions included.   
However, the nucleon and resonance kinetic energies are 
neglected in the later models, as in the FM, therefore their estimate
of the strength of $\Vtpip$ may be too large in magnitude. 
Another concern is that the $\pi N$ scattering amplitude used in these 
models considers only pions of momenta less than $m_{\pi}$~\cite{fhk99}. 
They play a much smaller role in nuclear binding than those with momenta 
$\sim 500$ MeV/c.  

The factor of $1/4$ in the second term of $\Otpip$ [Eq.(\ref{eq:otpip})],
containing the product of commutators, is due to the spin and isospin 
of the $\Delta$ being 3/2.  In the TM and later models the strength of this 
term is proportional to the constant $d$ whose values have also been 
tabulated by Friar et al.~\cite{fhk99}. The value of $d/b$ is 
0.29 in the latest Texas model, however, 
the ratio of the expectation values of the commutator and anticommutator
terms of $\Otpip$ is very constant across all the light nuclei studied in
this work, and hence this factor cannot be determined from the data 
considered here.
We continue to use the Fujita-Miyazawa value of 0.25 in this work for 
simplicity.

\subsection{$\Vtpis$}

The form of the $\Vtpis$, due to $\pi N$ S-wave scattering illustrated 
in Fig.~\ref{fig:vijk}b, in the TM model is:
\begin{equation}
B({\bf r}_{ij},{\bf r}_{jk})\ 
\left\{ \boldtau_i \cdot \boldtau_j , \boldtau_j \cdot \boldtau_k \right\} 
\left\{ ( S_{ij} + \boldsigma_i \cdot \boldsigma_j ) , ( S_{jk} + 
\boldsigma_j \cdot \boldsigma_k ) \right\} \ ,
\label{eq:tmsw}
\end{equation}
The $B({\bf r}_{ij},{\bf r}_{jk})$ contains several terms as given in 
Ref.~\cite{CPW83}.  We omit the short-range terms containing the 
$Z_0^{\prime}$ functions, whose validity has been questioned recently
\cite{fhk99}, and retain only 
the term with pion-exchange-range functions $Z_1^{\prime}$.  
These functions are given in Eqs.(A17,A18) of Ref.~\cite{CPW83}.
The function $Z_1^{\prime}$ is trivially related to the functions $Y(x)$ 
and $T(x)$ in $v^{\pi}_{ij}$ [Eqs.(\ref{eq:xij}-\ref{eq:vpi})], 
and the $Z_1^{\prime}$ contribution to $\Vtpis$ is expressed as:
\begin{eqnarray}
\Atpis &=& \left( \frac{f_{\pi NN}}{4 \pi} \right)^2 \ 
 a^{\prime} \ m^2_{\pi} \ , 
\label{eq:atpis}       \\
\Otpis &=&  \sum_{cyc} \  Z(m_{\pi}r_{ij}) Z(m_{\pi}r_{jk}) 
\boldsigma_i \cdot \hat{\bf r}_{ij} \boldsigma_k \cdot \hat{\bf r}_{kj}
\boldtau_i \cdot \boldtau_k   \ , 
\label{eq:otpis}       \\
Z(x) &=& \frac{x}{3}  \left[ Y(x) - T(x) \right]  \ 
 \ .
\label{eq:zij}
\end{eqnarray}
The values of the parameter $a^{\prime}$ are listed in Ref.~\cite{fhk99}; 
they vary from $-0.51/m_{\pi}$ to $-1.87/m_{\pi}$ in the recent models.  
The TM value $a^{\prime} = -1.03 / m_{\pi}$ gives $\Atpis \sim - 0.8$ MeV.
The $\Vtpis$ gives rather small contributions to nuclear energies, and it 
is difficult to extract its strength $\Atpis$ from nuclear data.  In model 
IL1 we neglect this term, while in all other models it is assumed to have 
the theoretically plausible strength of $-1$ MeV. 

\subsection{$\Vthpi$}

The present model of $\Othpi$ is based on the three-pion-exchange ring diagrams 
shown in Figs.~\ref{fig:vijk}c,d having only one $\Delta$ at a time 
in the intermediate
states.   The $\Vthpi$ is approximated with the sum of $\Vthpia$ and $\Vthpib$, 
which respectively denote the sums of diagrams \ref{fig:vijk}c and 
\ref{fig:vijk}d.
After neglecting the kinetic energies of the nucleons and the $\Delta$ in the 
intermediate states, diagram \ref{fig:vijk}c gives: 
\begin{equation}
V^{3\pi,\Delta R}_{1,ijk} = \sum_{cyc} \ \frac{1}{(m_{\Delta} - m_N)^2} \left[ 
v^{\pi}_{\Delta N \rightarrow NN}(ik)\ v^{\pi}(jk)\ v^{\pi}_{NN \rightarrow 
\Delta N}(ij) \ + \ j \rightleftharpoons k \right] \ ,
\label{eq:vthpia}
\end{equation}
where $v^{\pi}_{NN \rightarrow \Delta N}(ij)$, for example, 
denotes the one-pion-exchange 
transition potential~\cite{WSA84} exciting the nucleon $i$ to the 
$\Delta$-resonance state, and $ j \rightleftharpoons k $ denotes the term 
obtained by interchanging $j$ and $k$ in the previous term.  The above 
$\Vthpia$ can be reduced to a three-nucleon operator by eliminating 
the $N \rightleftharpoons \Delta $ transition spin and isospin operators 
denoted by ${\bf S}, {\bf S}^{\dagger}, {\bf T}$, and ${\bf T}^{\dagger}$
using the generalized Pauli identities:
\begin{eqnarray}
{\bf T}^{\dagger} \cdot {\bf T} &=& 2 \ ,    \\
{\bf T}^{\dagger} \times {\bf T} &=& - \case{2}{3}\ i\ \boldtau \ ,  \\
{\bf T}^{\dagger} \cdot {\bf A}\ {\bf T} \cdot {\bf B} &=& \case{2}{3}\ 
{\bf A} \cdot {\bf B} - \case{1}{3}\ i\ \boldtau \cdot {\bf A} \times 
{\bf B} \ ,
\label{eq:pitr}
\end{eqnarray}
for the transition isospin operators.  The transition spin operators also 
obey the same identities with $\boldtau$ replaced with $\boldsigma$. 
It is useful to reduce the $\Vthpia$ 
further by eliminating all the terms quadratic in either $\boldtau_l$ 
or $\boldsigma_l$ $(l = i,j,k)$ with the Pauli identity:
\begin{equation}
\boldsigma \cdot {\bf A}\ \boldsigma \cdot {\bf B} = {\bf A} \cdot 
{\bf B} + i\ \boldsigma \cdot {\bf A} \times {\bf B} \ , 
\label{eq:pi}
\end{equation}
for $\boldsigma$ and $\boldtau$ operators.  The resulting $\Vthpia$  
contains very many terms which can be organized in the following way:
\begin{eqnarray}
V^{3\pi,\Delta R}_{1,ijk} &=& \Athpi \ \Othpia \ , 
\label{eq:vao}  \\
\Athpi &=& \left( \frac{1}{3}\ \frac{f_{\pi NN}^2}{4 \pi}\ m_{\pi} 
\right)^3 \ \frac{f_{\pi N \Delta}^2}{f_{\pi NN}^2} \ \frac{1}
{(m_{\Delta}-m_N)^2} \ ,
\label{eq:athpi}   \\
\Othpia &=& 6\left(S^I_{\tau}S^I_{\sigma}+A^I_{\tau}A^I_{\sigma} \right)
+ 2 \sum_{cyc} \left( S^I_{\sigma}S^D_{\tau,ijk}+ S^I_{\tau}S^D_{\sigma,ijk}
+ A^I_{\tau}A^D_{\sigma,ijk}+ S^D_{\tau,ijk}S^D_{\sigma,ijk} \right) \ .
\label{eq:othpia}
\end{eqnarray}
The letters $S$ and $A$ denote operators that are symmetric and antisymmetric 
under the exchange of $j$ with $k$.  Subscripts $\tau$ and $\sigma$ label 
operators containing isospin and spin-space parts respectively, while 
superscripts $I$ and $D$ indicate operators that are independent or dependent 
on the cyclic permutation of $ijk$.  
The $S^I$ and $A^I$ would be more properly written with $ijk$ subscripts, but
because they are independent of the ordering of $ijk$, we omit them here for
brevity.
The isospin operators are:
\begin{eqnarray}
S^I_{\tau} &=& 2 + \case{2}{3}\left(\boldtau_i \cdot \boldtau_j 
+\boldtau_j \cdot \boldtau_k + \boldtau_k \cdot \boldtau_i \right) 
= 4 P_{T=3/2}  \ ,  \label{eq:sitau}  \\
A^I_{\tau} &=& \case{1}{3}\ i\ \boldtau_i \cdot \boldtau_j \times \boldtau_k 
= \case{1}{6} \left[ \boldtau_i \cdot \boldtau_j , \boldtau_j \cdot \boldtau_k \right]  \ ,
\label{eq:aitau}   \\
S^D_{\tau,ijk} &=& \case{2}{3} \boldtau_j \cdot \boldtau_k \ ,
\label{eq:sdtau}   \\
A^D_{\tau,ijk} &=& 0 \ ,
\end{eqnarray}
where we have indicated that $S^I_\tau$ is a projector onto isospin 3/2 triples 
(see the discussion at the end of this subsection), and that $A^I_\tau$ has
the same structure as the commutator part of $\Vtpip$.
The spin-space operators have many terms, and are 
listed in the Appendix.  In addition to the spin operators, they contain 
the functions $T(x)$ and $Y(x)$ in the $\vpi$.  The interaction 
$\Vthpia$ has to be symmetric under the exchange of $i$, $j$, and $k$;
therefore products of $S$- and $A$-type operators are not allowed.

The $\Vthpib$, obtained from diagram d in Fig.\ref{fig:vijk}, 
after neglecting the kinetic energies, is given by:
\begin{equation}
V^{3\pi,\Delta R}_{2,ijk} = \sum_{cyc} \ \frac{1}{(m_{\Delta} - m_N)^2} \left[ 
v^{\pi}_{N \Delta \rightarrow NN}(ik)\ v^{\pi}_{\Delta N \rightarrow N \Delta}
(jk)\ v^{\pi}_{NN \rightarrow N \Delta }(ij) \ + \ j \rightleftharpoons k \right] \ ,
\label{eq:vthpib}
\end{equation}
After appropriate reductions it can be cast in the form of $\Vthpia$ as 
follows:
\begin{eqnarray}
\Vthpib &=& \Athpib \ \Othpib \ , 
\label{eq:vbo}  \\
\Athpib &=& \Athpi \ \frac{f_{\pi N \Delta}^2}{f_{\pi NN}^2} \ ,
\label{eq:athpib}   \\
\Othpib &=& \case{8}{3}S^I_{\tau}S^I_{\sigma}+ \case{2}{3}A^I_{\tau}A^I_{\sigma}
-  \case{4}{9} \sum_{cyc} \left( S^I_{\sigma}S^D_{\tau,ijk}+ 
S^I_{\tau}S^D_{\sigma,ijk} + A^I_{\tau}A^D_{\sigma,ijk}- 
\case{1}{2}S^D_{\tau,ijk}S^D_{\sigma,ijk} \right) \ .
\label{eq:othpib}
\end{eqnarray}
The $\Vthpia$ and $\Vthpib$ are combined using $f^2_{\pi N \Delta} \sim 
4 f^2_{\pi NN}$ to obtain:
\begin{eqnarray}
V^{3\pi,\Delta R}_{ijk} &=& \Athpi \left( \Othpia + 4 \Othpib \right) = \Athpi \Othpi \ ,
\label{eq:vthpix}   \\
\Othpi &=& \case{50}{3}S^I_{\tau}S^I_{\sigma}+ \case{26}{3}A^I_{\tau}A^I_{\sigma}
\nonumber \\
&+& \case{2}{9} \sum_{cyc} \left(S^I_{\sigma}S^D_{\tau,ijk}+ 
S^I_{\tau}S^D_{\sigma,ijk} + A^I_{\tau}A^D_{\sigma,ijk}+
13~S^D_{\tau,ijk}S^D_{\sigma,ijk} \right)~.
\label{eq:othpix}
\end{eqnarray}
The strengths of the terms independent of cyclic permutations are larger than 
those which depend upon them.  Therefore we use the simpler $\Vthpi$ obtained 
by neglecting them, i.e. with the approximate operator:
\begin{equation}
\Othpi \approx \case{50}{3}S^I_{\tau}S^I_{\sigma}+ 
\case{26}{3}A^I_{\tau}A^I_{\sigma} \ .
\label{eq:othpi}
\end{equation}
The value of $\Athpi$ estimated from the observed values of the constants, 
and neglecting the kinetic energies, 
is $\sim 0.002$ MeV.  In all the Illinois models the $\Athpi$ is determined 
by fitting the nuclear energies. 

The $\Vthpi$ has an interesting dependence on the total isospin ${\bf T}_{tot}$ 
of the three interacting nucleons.  The $S^I_{\tau}$ can be written as:
\begin{eqnarray}
{\bf T}_{tot} &=& \case{1}{2}\left(\boldtau_i+\boldtau_j+\boldtau_k 
\right) \ , 
\label{eq:ttot}    \\
S^I_{\tau} &=& \case{4}{3}\ T^2_{tot} - 1 \ .
\label{eq:sttot}
\end{eqnarray}
Therefore the first term of $\Vthpi$ is zero in triplets having $T_{tot} = 1/2$,
i.e., in $d+N$ channels as well as in the $A=3,4$ bound states.  In 
contrast $A^I_{\tau}$ is zero in $T_{tot} = 3/2$ states.  It is therefore 
possible to extract the strength of this interaction from the data even though 
it is much weaker than the $\Vtpi$.  

\subsection{$\VR$}

The pion-exchange three-nucleon interactions are attractive, and  
lead to significant overbinding and large equilibrium density of nuclear 
matter.  Therefore there must be other three-nucleon interactions to 
compensate the attraction from $\Vtpi$ in nucleon matter at large density.  
In Faddeev calculations of the triton including $\Delta$ excitations~\cite{sauer,prb}, 
the attraction from processes included in $\Vtpi$ is more than cancelled by 
``dispersion'' terms which describe the modification of the contribution 
of the two-pion-exchange $\Delta$-box diagrams to $\vij$ due to the 
presence of the third nucleon $k$.  Such repulsive terms also occur in the 
variational theory in which $\Delta$-excitations are included via 
transition correlation operators~\cite{W84}.  The $\VR$ term in the Urbana 
models of $\Vijk$ was designed to approximate these effects.  It is retained in 
the Illinois models with the simple spin-isospin independent operator:
\begin{eqnarray}
O^R_{ijk} &=& \sum_{cyc} T^2(m_{\pi} r_{ij})\ T^2(m_{\pi} r_{jk}) \ ,
\label {eq:or}
\end{eqnarray}
Results of the Faddeev calculations~\cite{sauer,prb} indicate that the 
$\Delta$-effects not included in the $\Vtpi$ add $\sim$ 1 MeV to 
the energy of the triton.  The value of the 
strength $A_R$ required to obtain a contribution of $\sim$ 1 MeV from 
the $\VR$ to the triton energy is $\sim$ 0.004 MeV for a $T(r)$ with a
cutoff $c=2.1$~fm$^{-2}$.  In models IL1-4 
the $A_R$ is determined by fitting nuclear energies. 

\subsection{$\Vijk ^*$}

Most calculations of light nuclei use the simpler Hamiltonian obtained 
by approximating the $\vij$ in Eq.(\ref{eq:hnr}) by the $\tvij$.  The 
more accurate Hamiltonian: 
\begin{equation}
H^* = \sum_i - \frac{\hbar^2}{2m_i} \nabla_i^2 + \sum_{i < j} (\tvij + \dvij )
  + \sum_{i<j<k} \Vijk ^* \ ,
\label{eq:hnrst}
\end{equation}
contains the boost correction to the two-nucleon interaction. 
This correction $\dvij$ is of first order in $P^2_{ij}$, and therefore its 
contribution is calculated as a first-order perturbation
in wave functions generated by a nonrelativistic Hamiltonian
that gives the same final energy as $H^*$.  
In light 
nuclei the expectation values of $\dv$ and $\VR$ are nearly proportional 
to each other~\cite{rel1,rel2}.  Therefore the energies of light nuclei, 
calculated with the simpler $H$ can be reproduced with the $H^*$ using:
\begin{equation}
\Vijk^* = \Atpip \Otpip + \Atpis \Otpis + \Athpi \Othpi + \AR^* \OR \ .
\label{eq:vijkst}
\end{equation}
which differs from $\Vijk$ only in the strength of $\VR$.  For $T(r)$ with 
cutoff $c=2.1$ fm$^{-2}$ the 
strength $A_R^*$ in $\Vijk^*$ is smaller than $A_R$ by $\sim$ 0.002 MeV.  
The proportionality of $\dv$ and $\VR$ contributions appears only in bound 
light nuclei.  In nuclear
matter the $\dv$ contribution increases more slowly with density than that of  
$\VR$, and in neutron drops the $\dv$ gives a relatively larger contribution. 
Therefore it is necessary to use the $H^*$~\cite{APR98} in these systems.  

\subsection{The Illinois Model-5}

This model is meant to test the sensitivity of nuclear energies to the 
spatial shape of $\VR$.  Here we assume that:
\begin{equation}
V^R_{ijk} = 0.002\ \OR + A_W \prod_{cyc} W(r_{ij}) \ ,
\end{equation}
where $W(r)$ is a modified Woods-Saxon function that has zero-derivative
at the origin: 
\begin{eqnarray}
W(r) &=& \frac{1}{1 + e^{(r-r_W)/a_W}}  
      \left[ 1 + \frac{r/a_W}{1 + e^{r_W/a_W}} \right]  ~,
\end{eqnarray}
with $r_W = 1.0$ fm and $a_W = 0.2$ fm.
The first term of this $\VR$ 
is meant to take into account the contribution of $\dv$ 
omitted from the simpler $H$, and the second corresponds to the 
phenomenological three-nucleon repulsion. 

\subsection{Potential Parameters}

The parameters of the Illinois potentials were determined by fitting 
the observed energies of $3~\leq~A~\leq~8$ nuclei. 
The simpler Hamiltonian $H$ was used, and a 
total of 17 ground or excited states with widths less than 200 
keV were considered.  Table~\ref{table:params} presents the parameters
of the new potentials, and for comparison those of UIX~\cite{PPCPW97}.  
The properties of light nuclei calculated from Hamiltonians including
the AV18 $\vij$ and the new $\Vijk$ are presented below in Sec. V.
In addition to the various strengths, Table~\ref{table:params} gives the
value, $c$, of the cutoff constant in Eq.(\ref{eq:cutoff}) which is
used in all the four terms of $V_{ijk}$. 

As mentioned earlier, this data set cannot determine all five 
parameters of the Illinois $V_{ijk}$.  At most three parameters were 
varied for each model,  and plausible values are assumed for 
the others.  These assumed values are marked with an asterisk
in Table \ref{table:params}.  There is a substantial cancellation 
between the contributions of $\Vtpip$ and $\VR$ in the nuclear 
binding energies.  Therefore one can make correlated changes in 
the $\Atpip$ and $\AR$, as in models IL2 and IL4, 
without significantly spoiling the fit. 
Presumably nuclear matter calculations with these models 
can help to further constrain the parameters. 

As will be discussed later, the contributions of $\Vtpis$ and
$\Vtpip$ are in a fairly constant ratio for the light nuclei considered
here, and thus we cannot uniquely determine both $\Atpis$ and $\Atpip$. 
IL1 assumes that $\Atpis=0$, while in all other models $\Atpis=-1$ MeV 
as in modern chiral-perturbation-theory potentials~\cite{fhk99}.  
The $\Atpip$ in IL2 is less than that in IL1 by $\sim 4\%$ to 
compensate for the $\Vtpis$ contribution. 
All models other than IL3 have the same cutoff, $c$, as in
\avet and UIX and have $\Atpip$ of the same order
of magnitude as the Urbana models.  This $\Atpip$ is approximately
half that favored by chiral-perturbation theory.  We constructed the
IL3 to see if light nuclei are sensitive to this difference.
In this model, $\Atpip$ was fixed at a typical chiral-perturbation theory
value and the cutoff parameter, $c$, and strengths $\AR$ and $\Athpi$ 
were adjusted to fit the binding energies.  The cutoff had to be
made much softer to compensate for the strong $\Atpip$. 

The expectation values of $\dv$ were calculated in a few nuclei for each 
model, and the values of $A^*_R$ were estimated requiring: 
\begin{equation}
\langle \Vijk - \Vijk^* \rangle \approx \langle \dvij \rangle \ . 
\label{eq:bdiff}
\end{equation}
They are listed in Table \ref{table:params}.  

\section{Quantum Monte Carlo Calculations}

The GFMC calculations presented below were made using essentially
the same methods and variational wave functions as described in
Ref.~\cite{WPCP00} for our
calculations of $A \leq 8$ nuclei with the UIX three-nucleon
potential.  Here we describe only the few enhancements that were
necessary for using the Illinois three-nucleon potentials.

The new terms in the Illinois potentials are static and hence present no
formal difficulties for the GFMC propagator beyond those already
encountered for the UIX $\Vijk$: they are included by expanding
$\exp(- \case{1}{2} \Vijk \Delta \tau)$ to first order, as in Eq.(4.5) of
Ref~\cite{WPCP00}.  The structure of $\Vtpis$ is similar to that of
the anticommutator part of $\Vtpip$ ($V^{2\pi,A}$) and can also be reduced to just
two-body operators in spin-isospin space.  Thus it can be combined
with the $\alpha V^{2\pi,A}_{ij;k}$ in Eq.(4.21) of Ref~\cite{WPCP00}
with almost no increase in the required computer time.  

The $\Vthpi$ involves many operators which are unfortunately not
reducible to two-body operators in spin-isospin space, and thus, like
$V^{2\pi,C}$, the commutator term of $\Vtpip$, it adds a lot of time to the
evaluation of a propagation step.  In Ref~\cite{WPCP00}, we showed that
$V^{2\pi,C}$ could be replaced with an increased strength of
$V^{2\pi,A}$ in the propagator with no loss of accuracy.  The $\Vthpi$
cannot be similarly completely replaced, but one can use two to four
steps of propagation with just an enhanced $V^{2\pi,A}$ and then a
correction; see Eq.(4.23) of Ref~\cite{WPCP00}.  In general the
tests we have made show that systematic errors in the GFMC calculations
are less than 1\% for the total energy; however $^8$He appears to be
a particularly difficult case and the systematic errors for it are
probably 2\%.

The GFMC calculations for the models with $\Vijk$ and p-shell nuclei were 
carried out as described in Ref.~\cite{WPCP00}. 
In particular, propagations were made to $\tau=0.2$~MeV$^{-1}$ with steps of
$\Delta\tau=0.0005$~MeV$^{-1}$ (400 steps) and expectation values were
computed every 20 steps with averages of the last 7 values ($\tau\geq
0.14$~MeV$^{-1}$) being used.  The s-shell calculations for most models
and some of the p-shell calculations for models without $\Vijk$ were propagated
to only $\tau=0.1$~MeV$^{-1}$.  In most cases 10 unconstrained steps
were used before each energy evaluation [$n_u=10$; see Eq.(4.17) of 
Ref.~\cite{WPCP00}], but 20 steps were used for some cases.

The $^4$He energy obtained 
with the Yakubovsky equations~\cite{NKG00} and AV18/UIX Hamiltonian,
$-28.50(5)$ MeV, is within $\sim 0.5\%$ of the latest GFMC~\cite{WPCP00} 
result, $-28.33(2)$ MeV.  The GFMC calculations are carried out with an 
approximation called AV8$^{\prime}$~\cite{PPCPW97}, containing only the first 8 
operators given in equations (\ref{eq:stos}) and (\ref{eq:soos}). 
The small difference between the AV18 and AV8$^{\prime}$ is treated 
perturbatively.  It is therefore likely that the exact results for 
AV18/UIX are a little below the present GFMC results.  The differences
between GFMC and other calculations of binding energies of three and four
nucleons will be the subject of another paper.

\section{Results - Light Nuclei}

This section presents various results for $3{\leq}A{\leq}8$ nuclei and neutron 
drops using the new Illinois models, the AV18/UIX model, and Hamiltonians
containing just the \avep or \avet potentials with no $\Vijk$.  
All of these results were obtained from GFMC calculations.  
With the exception of the total energy, which is discussed in the next 
paragraph, the values are perturbatively extrapolated from the mixed estimates,
as described in Ref.~\cite{PPCPW97}.  
Monte Carlo statistical errors are given in parentheses, but no estimate is
made for the systematic errors associated with the extrapolation of mixed 
expectation values.  The results presented in subsections A to F are with 
the simple $H$ without boost interaction, whose contributions are 
reported in subsection G. 

Most experimental energies and moments are drawn from the standard
compilations of Ajzenberg-Selove~\cite{AS88} and the TUNL Nuclear Data 
Evaluation Project~\cite{TUNL3,TUNL4,TUNL567}, while
charge radii are taken from the NIKHEF compilation~\cite{NIKHEF87}.
More recent data not included in these references are the energy of the 
$^8$He(2$^+$) state~\cite{he8-2plus}, the charge radius of $^3$He~\cite{SDV95}, 
and the $A=8$ magnetic and quadrupole moments~\cite{R89}.

\subsection{Energies of ``Narrow'' States}

Table~\ref{table:narrow-energies} shows our GFMC energies for
$3{\leq}A{\leq}8$ ``narrow'' states for all the Hamiltonians along with
the corresponding experimental energies.  
The states are either particle stable, or have experimental widths less than
200 keV, and are used to fit the parameters of the Illinois
models.  The \avep and \avet Hamiltonians have no three-nucleon
potential; they are presented to show the importance of the $\Vijk$, and
to provide results for comparison with those from other many-body
methods.  

The \avep model consists of the Argonne $\v8p$ two-nucleon potential,
which is an eight-operator refit of the \avet, and the isoscalar Coulomb
potential, as defined in Ref.~\cite{PPCPW97}.
The \avep results should be the most reliable of all
our results, because the GFMC propagation is made with the same
Hamiltonian as is used for the energy expectation values, as
discussed in Ref.~\cite{PPCPW97}.  In all other cases an effective $H'$
is used for propagation and a small contribution $\langle H-H' \rangle$
is evaluated perturbatively.  In the cases that have a
three-nucleon potential, the $\AR$ is adjusted in $H'$ to make
$\langle~H-H'~\rangle~\sim~0$; however such an ability does not exist
for the \avet with no $\Vijk$.  The \avet energies are found by
perturbatively evaluating $\langle\vet-\v8p\rangle$ in the \avep calculation.

As described in the previous section, the
AV18/IL1 and AV18/IL2 models differ only in the values of $\Atpis$ and
$\Atpip$, such that $\langle~\Vtpis+\Vtpip~\rangle$ in IL2 is nearly the same
as $\langle~\Vtpip~\rangle$ in IL1.  We made a complete set of calculations for
AV18/IL2 and perturbatively computed $\langle~\Vtpip~\rangle$
for AV18/IL1 in the AV18/IL2 wave functions.  This result was used
to generate 12 of the 21 AV18/IL1 energies in Table~\ref{table:narrow-energies}. 
The procedure was
checked by making a new propagation using the AV18/IL1 $H^\prime$, and these
explicitly calculated values are shown for nine states; they are not significantly
different from the perturbative estimates,
i.e., the differences are generally smaller than the Monte Carlo errors.

The five Illinois Hamiltonians give very similar energies.  The
predictions for the p-shell nuclei are significantly better than those
obtained with the AV18/UIX model.  
This is illustrated in Fig.~\ref{fig:energies} where results for the
AV18/UIX, AV18/IL2, and AV18/IL4 models are compared to experiment.
The relative stability of the helium and lithium nuclei with the Illinois
models is clearly evident, as is the just unbound nature of $^8$Be.
More quantitatively, Tables~\ref{table:energy-errors} and
\ref{table:excitation-errors} show various averages of the deviations from
experiment for the narrow states of Table~\ref{table:narrow-energies}.
Table~\ref{table:energy-errors} is based on the deviations of the total
energies of the 17 states, while Table~\ref{table:excitation-errors} is
based on the deviations of the excitation energies of excited states.  Both
tables show the average deviation (which includes the signs of the
deviations), the average of the magnitudes of the deviations, and the rms
deviations.  
The average deviations in Table~\ref{table:energy-errors} demonstrate that the
Hamiltonians with no $\Vijk$ systematically underbind these nuclei by 5
to 7 MeV; AV18/UIX reduces this to the still large value of 2 MeV
underbinding.  The five Illinois models have no significant systematic
under or overbinding.  Because the errors for the
\avep, \avet, and AV18/UIX cases are so one-sided,
their average absolute and rms errors are comparable to their average
signed errors.  The rms error obtained with the AV18/IL1-5 Hamiltonians 
is $\sim$ 1\%. 

Table~\ref{table:excitation-errors} shows that the Hamiltonians with no
$\Vijk$ produce a spectrum that is too compressed, although it has far
smaller deviations than the absolute energies.  All of the
Hamiltonians with $\Vijk$ produce excitation
spectra with significantly smaller rms deviations.  However some of 
them (particularly AV18/IL2) are too expanded.
The excitation spectra obtained with
AV18/IL3 and AV18/IL4 appear to be somewhat better than the others.

\subsection{Contributions to the Energies}

The contributions of the two-body potentials, $\vij$, including electromagnetic
terms, and the sum ($K+\vij$) are shown
for several Hamiltonians in Table~\ref{table:vij}.  As was discussed in
Ref.~\cite{PPCPW97}, the perturbatively extrapolated values of
$\langle{K}\rangle$, $\langle{\vij}\rangle$, and $\langle{\Vijk}\rangle$
do not add up to the total energy, $\langle{H}\rangle$.  The latter is
the most reliably computed quantity.  Other studies of GFMC calculations
\cite{GFMC-extrap} suggest that the perturbative extrapolation of the
potential energy is more reliable than that of the kinetic energy.
Therefore the values of $\langle{K+\vij}\rangle$ are obtained by
subtracting $\langle{\Vijk}\rangle$ from $\langle{H}\rangle$; estimates
of $\langle{K}\rangle$ may be obtained by subtracting $\langle{\vij}\rangle$
from $\langle{K+\vij}\rangle$.

Table~\ref{table:vij} shows some of the non-perturbative aspects of
these calculations.  For the p-shell nuclei, the total binding energy
steadily increases from \avet to AV18/UIX to the Illinois
models.  However the values of $\langle{K+\vij}\rangle$, which involve
the same operators in all cases, steadily decrease in magnitude.  This is because the
wave function is being tuned to the ever stronger $\Vijk$ and hence is
becoming less favorable for $K+\vij$.  The net increase in the binding
energy comes from even bigger increases in $\langle{\Vijk}\rangle$ (see
Table~\ref{table:Vijk}).  Although $\langle{K+\vij}\rangle$ is becoming
less attractive in this progression, $\langle{\vij}\rangle$ is becoming
more negative due to the enhanced tensor-isospin (pion-exchange)
correlations induced in the wave function by $\Vijk$; these correlations
also increase the kinetic energy.

Table~\ref{table:Vijk} shows the total $\langle{\Vijk}\rangle$ for the
various models.  The AV18/IL1 and AV18/IL2 models were constructed to have
approximately the same $\langle{\Vijk}\rangle$.  Although the AV18/IL3
model has very similar total binding energies as AV18/IL1 and AV18/IL2,
there are significant differences in many of its contributions.
This indicates that the correlations induced
by IL3 (which is stronger and has a much softer pion form factor) in the wave
functions make important non-perturbative changes in
$\langle{K+\vij}\rangle$.  
The AV18/IL4 and AV18/IL5 models have weaker strengths than
the first three and smaller net $\langle{\Vijk}\rangle$.
As expected, the $\langle{\Vijk}\rangle$ for
the Illinois models are all larger than for the AV18/UIX model in
p-shell nuclei; they are also larger for the s-shell nuclei even though
all the models give the same binding energies for these nuclei.

The fraction of the total binding energy represented by
$\langle{\Vijk}\rangle$ increases from s- to p-shell nuclei and as $N-Z$
increases. For AV18/IL2 it is 17\% for $^3$H, 30\% for $^4$He, but then a
nearly constant 33-37\% for $^{6,7}$Li and $^8$Be. It then jumps to
49\% for $^8$Li and 52\% for $^8$He.  Expressed as a
fraction of the total potential energy, the AV18/IL2 $\langle{\Vijk}\rangle$
ranges from 2.5\% for $^3$H to 6.1\% for $^4$He up to 7.5\% for $^8$He.
These fractions are typical of the other Illinois models except that AV18/IL3
has somewhat larger ratios.

The individual contributions to $\langle{\Vijk}\rangle$ are shown in
Tables~\ref{table:IL2-terms} and \ref{table:IL3-terms} for AV18/IL2 and AV18/IL3,
respectively.  The ratios of the contributions of the anticommutator and
commutator parts of $\Vtpip$ [see Eq.(\ref{eq:otpip})] for the nuclei studied
here are remarkably constant for a given model.  For AV18/IL2 the ratio of the
anticommutator contribution to the total $\langle{\Vtpip}\rangle$ varies only
between 0.62 and 0.63 for the nuclei in Table~\ref{table:IL2-terms}, 
while it is 1.0 for pure
neutron systems (the $\left[\boldtau_i\cdot\boldtau_j,
\boldtau_j\cdot\boldtau_k\right]$ is zero in $T=\case{3}{2}$ triples).  
Very similar ratios are obtained for AV18/UIX and the other Illinois models.  
A similar ratio has also been found
in VMC calculations of $^{16}$O using an older Urbana model~\cite{PWP92}.  
It is because of this very small variation
that one cannot improve fits to the energies of light nuclei
by changing the factor of $\case{1}{4}$ in Eq.(\ref{eq:otpip}).

The ratio of the contribution of $\Vtpis$ to that of $\Vtpip$ is also quite
independent of nucleus.  For AV18/IL2 it ranges from 3.6\% for $^8$He to 4.2\% 
for $^3$H. This ratio depends upon the model; for example, it is $\sim$ 2.3\%  
for AV18/IL3 and $\sim$ 5\% for AV18/IL4. Thus 
small changes in $\Atpip$ can accurately compensate large changes in $\Atpis$,
making it impossible to uniquely determine the value of  $\Atpis$ from the 
binding energies of light nuclei.  We have made versions
of AV18/IL2 and AV18/IL3 that have the unreasonably large value of 
$\Atpis=-2.2$ MeV; these also gave good fits to the binding energies of light nuclei.  

The $S^I_{\tau}S^I_{\sigma}$ and $A^I_{\tau}A^I_{\sigma}$ terms of $\Vthpi$ 
are shown in the columns labeled $V^{3\pi,SS}$ and $V^{3\pi,AA}$, respectively.
The $S^I_{\tau}$ is a projector onto $T=\case{3}{2}$ triples and thus the 
$S^I_{\tau}S^I_{\sigma}$
term of Eq.(\ref{eq:othpi}) vanishes in s-shell nuclei.  The $A^I_{\tau}A^I_{\sigma}$ 
term results in
a repulsive contribution in nuclei and hence $\Vthpi$ is repulsive in s-shell
nuclei.  In p-shell nuclei the $S^I_{\tau}S^I_{\sigma}$ term of $\Vthpi$ is
attractive and larger in magnitude than the $A^I_{\tau}A^I_{\sigma}$ term.
Thus the $\Vthpi$ changes sign between s-shell and p-shell nuclei and also
becomes more attractive as $N-Z$ increases.  This allows $\Vthpi$ to substantially
improve the fit to the energies of light nuclei.  Its strength is therefore 
rather well determined by the data in the context of present models.  
In low-density neutron drops the $\Vtpi$ terms become very small, while the
$\Vthpi$ is attractive and gives the largest contribution to $\Vijk$.

An interesting
property of the $A^I_{\tau}A^I_{\sigma}$ term is that it only increases
by a factor of $\sim 3$ as $A$ increases from 3 to 4, while all the other 
$\Vijk$ terms rise by a factor of 5.5.  
The factor of 5.5 can be understood from the simple argument that $^4$He has
four triples compared to only one triple in the trinucleons, and each triple 
in $^4$He gives $\sim$ 40\% more contribution due to the higher density from 
increased binding.  The $\Vtpi$ and $V^R$ have relatively simple spatial 
dependence given by a product two radial functions; in contrast the 
$A^I_{\sigma}$ has a more complex spatial dependence, given by 
Eq.(\ref{eq:AIsig}), 
containing radial functions of all the three pair distances. 
In principle, the strength of the $A^I_{\tau}A^I_{\sigma}$ term in $\Vijk$ 
can be adjusted separately from the $S^I_{\tau}S^I_{\sigma}$ term to reproduce 
the energies of $^4$He and the trinucleons 
with the desired precision; however, agreement with less 
than 1\% error can be obtained assuming that the strengths of the 
$S^I_{\tau}S^I_{\sigma}$ and $A^I_{\tau}A^I_{\sigma}$ terms of $\Vthpi$ 
have the theoretical ratio of $50/26$ from Eq.(\ref{eq:othpi}). 

Table~\ref{table:IL2-vpi-r} shows the $\vpi$, $\Vtpi$, $\Vthpi$, and the 
remainder contributions to the potential-energy expectation values, 
evaluated for AV18/IL2.  Note that additional pion-exchange contributions 
omitted from $\vpi$, $\Vtpi$ and $\Vthpi$ are contained in the $v^R$ 
and $V^R$.  For example, the two-pion-exchange two-nucleon interaction 
provides most of the intermediate-range part of $v^R$.  
The total two-pion terms of $\Vijk$ are typically
11\% of the one-pion part of $\vij$ for the AV18/UIX, AV18/IL4, and AV18/IL5 models,
while they are 16\% for AV18/IL2 and 21\% for AV18/IL3.  
The ratio of $\Vthpi$ to $\Vtpi$ 
changes sign between s-shell and p-shell nuclei.

The ratio of the $\Vtpi$ and $\VR$ contributions for a given model does not
change by much in the 
light nuclei.  For the AV18/IL2 model, the ratio is $-2.23$ for
s-shell nuclei and $-$1.97 for $^8$Li.  The AV18/IL4 model
has the largest range, from $-$2.8 for s-shell nuclei to $-$2.35 for $^8$Li.
When only $^3$H and $^4$He are included in the fit, as in the case of the UIX 
model, it is not possible to determine $\Atpip$ and $\AR$ separately. 
For this reason the equilibrium density of nuclear matter was used in 
Ref.~\cite{PPCPW97} to determine $\AR$ even though exact calculations of 
nuclear matter properties are not yet possible.  

Even after including all $A \leq 8$ 
nuclei in the fit, we cannot determine $\Atpip$, $\AR$ and $\Athpi$ separately. 
For example, in AV18/IL2, decreasing $\AR$ and $\Atpip$ by factors of 0.5 and 0.77 
leaves the energies of $^3$H and $^4$He unchanged in first order, however, 
that of $^8$Li decreases by 0.8 MeV, {\em i.e.} by $\sim 2\%$.  This 
change can be compensated by reducing $\Athpi$ by $\sim 20\%$.  Thus in 
first order multiplying $\AR$, $\Atpip$ and $\Athpi$ by 0.5, 0.77 and 0.8 
leave the energies of s-shell nuclei and $^8$Li unchanged. 

The strengths of $\AR$, $\Atpip$ and $\Athpi$ in IL4 are respectively 
0.55, 0.76 and 0.81 times those in IL2.  The difference between these 
ratios and those in the preceding paragraph is due to nonperturbative effects. 
The overall fit obtained with IL4 is slightly better than that with IL2 
(see Tables~\ref{table:energy-errors} and~\ref{table:excitation-errors}),
especially for excitation energies, 
suggesting that the parameters of IL4 have more realistic values. 
Preliminary results obtained for $A=9,10$ nuclei also indicate that 
IL4 is better than IL2.  

\subsection{Isobaric Analog Energies}

Table~\ref{table:isoenergies} gives the total isovector $(n=1)$ 
and isotensor $(n=2)$ 
energy coefficients, $a^{(n)}_{A,T}$, defined in Eq.(5.3) of
Ref.~\cite{WPCP00}, for energy differences in isobaric multiplets.
The first three lines of the table show three different evaluations
of the $a^1_{3,\case{1}{2}}$.  The first two are expectation values
of the isovector operators in separately computed $^3$H and $^3$He 
GFMC wave functions, respectively.  The last is the difference of
separately computed GFMC energies; it has a considerably larger
statistical error, but otherwise would be the best calculation to
compare to experiment.  The expectation values computed in the
two different wave functions are statistically different, but, due
to its larger error, the energy difference is consistent with both. 

The $a^{(n)}_{A,T}$ in multiplets having $A>3$ (Table~\ref{table:isoenergies}) 
have been computed as expectation values 
of the isovector and isotensor parts of the Hamiltonians 
in the wave function for the $Z=\case{1}{2}A-T$ nucleus.
Table~\ref{table:IL2-isoenergies} gives a breakdown of these
coefficients into various contributions for the AV18/IL2 model.
These include: $K^{CSB}$, the kinetic energy contribution due to the
neutron-proton mass difference; $v_{C1}(pp)$, the proton-proton Coulomb
term; $v^{\gamma,R}$, the remaining electromagnetic contributions, such as
magnetic moment interactions, which are part of Argonne $\vet$; 
and the strong-interaction terms $v^{CSB}$ and $v^{CD}$.  
The isotensor $v^{CD}$ comes
from components 15$-$17 of Argonne $\vet$ and contributes only to
the $a^{(2)}$, while the isovector 
$v^{CSB}$ is term 18 and contributes only to $a^{(1)}$.

The isovector terms are dominated by $v_{C1}(pp)$, the expectation value
of which is strongly correlated with the rms radii.
The Illinois models generally give better total binding energies and radii
(see below) and thus better values for these coefficients.
However, the remaining kinetic and potential terms contribute
5--10\% of the total isovector energy coefficient, thus playing an important
part in the overall good agreement with experiment.
The isotensor terms are also dominated by $v_{C1}(pp)$, but in this case
the increased binding of the Illinois models has not improved the 
agreement with experiment.  The $v_{C1}(pp)$ alone underestimates the 
$a^{(2)}_{A,T}$; the strong interaction contributions have the correct 
sign, but seem to be too large in magnitude.

\subsection{Energies of ``Wide'' States and Spin-orbit Splittings}

So far we have presented results for states of nuclei that are either
particle stable or have narrow experimental widths.  Our GFMC
calculations which treat all states as bound systems should be reliable
for such states, and the comparison of the resulting energies to
experimental values should be unambiguous.
Table~\ref{table:other-energies} shows GFMC calculations of additional
states that are experimentally broad or experimentally unknown.  An
example of the GFMC propagation for broad and narrow states was shown in
\cite{WPCP00}; the energy of the broad state falls slowly but steadily
with imaginary time.  This introduces a certain, but usually small,
ambiguity in determining the resonance energy from the calculation; we
use the average of the the energies for
0.14$\leq\tau\leq$0.20~MeV$^{-1}$.  In addition, the experimental
assignment of some resonance energies may be difficult.  Nonetheless,
the rms errors in the Illinois predictions of the experimentally known
states in Table~\ref{table:other-energies} are only $\sim$700~keV.

It has long been known that Hamiltonians containing only realistic
two-nucleon potentials often cannot correctly predict the observed spin-orbit
splitting of nuclear levels; in fact one of the original motivations
for the Fujita-Miyazawa three-nucleon potential was the study
of spin-orbit splittings~\cite{FM57}.  In~\cite{PP93} we showed that
one of the Urbana family of $\Vijk$ makes a substantial contribution
to the spin-orbit splitting in $^{15}$N.  Table~\ref{table:so-splits}
shows calculated and experimental splittings for a number of states
that are spin-orbit partners in conventional shell-model calculations.
The dominant $L$ and $S$ in the shell-model calculations (and in
the one-body parts of our variational wave functions) are shown.

The spin-orbit splitting computed with just the two-nucleon interactions,
\avep or \avet are generally too small, sometimes by factors of two to three.
In some cases AV18/UIX makes a significant increase, but in general
it also predicts too small splittings.  The predictions with the Illinois
models are in much better agreement with the experimental values.
Due to significant statistical errors in the calculated spin-orbit 
splittings and the fact that some of the spin-orbit partners are wide
states, they cannot yet be used to differentiate between the various 
Illinois models. 

\subsection{Point Nucleon Radii and Electromagnetic Moments}

Table \ref{table:pn-radii} gives the point proton
and neutron radii for some of these models.  The ``experimental'' point proton
radii were obtained by subtracting a proton mean-square radius of 0.743 fm$^2$
and $\case{N}{Z}$ times a neutron mean-square radius of --0.116 fm$^2$
from the squares of the measured charge radii.  As mentioned earlier 
the GFMC propagations are carried out for an $H'$, and results for 
the desired $H$ are obtained by treating $H-H'$ as a first order 
perturbation.  Therefore the point radii and electromagnetic moments are
computed for $H'$ instead of $H$.  For the models with three-body
potentials, the $H'$ has a modified $A_R$ such that $\langle~H-H'~\rangle~\sim~0$, 
and thus the radii and moments for the $H'$ should be close
to the desired ones for $H$.  However the calculations for \avet with
no three-body potential use just \avep for $H'$.  In this case $\langle~H-H'~\rangle$
is significantly different from zero, and we can only quote radii and moments
for \avep.
For the few light p-shell nuclei that have measured radii, those obtained
with the Illinois models, which produce good binding energies, are in
better agreement with the data than those obtained with either 
\avep or AV18/UIX.

Table \ref{table:magmom} shows the experimental
isoscalar and isovector magnetic moments for the cases that have been
measured along with values calculated using only one-body
current operators.  The values in the table are defined in the same way as the
coefficients $a^{(n)}_{A,T}$ in Table~\ref{table:isoenergies} (see
Eq.(5.3) of Ref.~\cite{WPCP00}) and thus the isovector $(\mu^{(1)})$ values for
$T=\case{1}{2}$ cases are $-$2 times those often quoted.  The $\mu^{(0)}$ 
and $\mu^{(1)}$ are obtained from expectation values of the 
isoscalar and isovector magnetic-moment operators in the 
wave function for the nucleus having smallest $Z=\case{1}{2}A-T$. 
In this approximation the isotensor $\mu^{(2)}=0$, since the one-body
magnetic-moment operator does not have an isotensor term.  However, one 
may obtain a small $\mu^{(2)}$ when the magnetic moments are separately calculated 
for each state in the multiplet due to violation of isospin symmetry 
in the wave functions. 

For the $A=8$, $T=1$ nuclei, the experimental isoscalar and isovector moments 
are obtained from the sum and difference of the values for B and Li, since the 
magnetic moment of the $T=1$ $J^{\pi}=2^+$ state in Be is not measured. 
In fact the sum gives $2\mu^{(0)}+\mu^{(2)}$, but in the present 
approximation $\mu^{(2)}=0$. 

The computed magnetic moments show little dependence on the three-nucleon interaction.
Because pair currents are not included in the calculated values
in Table \ref{table:magmom}, one cannot expect good agreement with
the experimental values, especially for the isovector values.  The 
pair-current corrections computed for the
$A=3$ system using the AV18/UIX model are 0.034 and $-$0.778 for
the isoscalar and isovector moments, respectively~\cite{MRS98}.
This isoscalar correction is twice what is needed to achieve 
agreement with experiment while the isovector value results in
perfect agreement.  All the computed magnetic
moments differ from the experimental values by amounts comparable
to the corrections computed for $A=3$.  

Table \ref{table:quadrupole} shows computed (again using just
impulse approximation) and experimental quadrupole moments.
The Illinois models predict quadrupole moments that are generally smaller
than those obtained using AV18/UIX.  This is a consequence
of the increased binding energy, and resulting smaller rms radii.
The situation for $^6$Li is very difficult due to the large cancellation
between orbital and intrinsic components (in an alpha-deuteron picture)
of the quadrupole moment.

\subsection{Neutron Drops}

Neutron drops are systems of interacting neutrons confined in an
artificial external well.  We have previously reported results for
systems of seven and eight neutrons as a basis for comparing Skyrme
models of neutron-rich systems with microscopic calculations based on
realistic interactions~\cite{PSCPPR96}.  The determination of the
isospin dependence of the Skyrme model spin-orbit parameters is of
particular interest.  The external one-body well that we use is a
Woods-Saxon:
\begin{equation}
V_1 (r) = \sum_i \frac{V_0}{1 + \exp [ - (r_i - r_0)/a_0 ]} \ ;
\end{equation}
the parameters are $V_0 = -20$ MeV, $r_0 = 3.0$ fm, and $a_0 = 0.65$ fm.
Neither the external well nor the total internal potential
($\vij+\Vijk$) are individually attractive enough to produce bound
states of seven or eight neutrons; however the combination does produce
binding.

Many of the tables show results for the neutron drops.  
The $T=\case{3}{2}$ nature of the
$S^I_{\tau}S^I_{\sigma}$ term of $\Vthpi$ results in large contributions
in the neutron drops.  As a result the seven-neutron drops computed with
some of the Illinois potentials have double the spin-orbit splitting predicted by
AV18/UIX.  This strong dependence on the Hamiltonian
indicates that the conclusion of
\cite{PSCPPR96}, that conventional Skyrme models over-predict the spin-orbit
splitting in neutron-rich systems may not be valid.  The IL3 model gives 
larger spin-orbit splitting in the $A=7$ neutron drop than the others. 

\subsection{$\dv$ Contributions}

Table \ref{table:dv} shows the expectation values of $\dv$ in various nuclei 
for some of the Illinois Hamiltonians along with the net change in the 
binding energy due to the boost correction: 
\begin{equation}
\langle H^* - H \rangle = \langle \dv + (\Vijk^*-\Vijk) \rangle ~ .
\end{equation}
In the light 
nuclei the net change is, at most, comparable to 
1\% of the binding energies, and therefore the Hamiltonian $H^*$ 
[Eq.(\ref{eq:hnrst})] gives essentially the 
same energies as the simpler $H$ [Eq.(\ref{eq:hnr})] without $\dv$ correction. 
However, the net change is not necessarily small in other nuclear 
systems like nuclear or neutron matter or neutron drops.  In the $A=8$ 
drop (Table~\ref{table:dv}) the $\langle \Vijk-\Vijk^* \rangle$ is only 
half as large as $\langle \dv \rangle$. 
In these systems we must use the Hamiltonian $H^*$. 

\section{Conclusions and Discussion}

We find that nonrelativistic Hamiltonians containing two- and three-nucleon 
interactions can reproduce the energies of all the bound and narrow 
states of up to eight nucleons with an rms error $\alt$ 1\% via GFMC 
calculations, which have an estimated error of $<$ 2\%.  The three-nucleon 
interactions give a significant fraction of nuclear binding energy 
due to a large cancellation between the kinetic and two-nucleon 
interaction energies.  This cancellation is seen even in the deuteron 
whose kinetic and interaction energies obtained from AV18 are 
respectively $+ 19.9$ and $- 22.1$ MeV.  Since the AV18 model is 
very successful in explaining the observed deuteron form factors, 
and all realistic models of $\vij$ also have this feature, 
it seems to be inherent in nuclei. 

The Bochum group~\cite{NHKG97} obtained a one
parameter family of $\Vijk$ models by choosing the cutoff mass in the
TM model of $\Vtpi$, including P- and S-wave terms, to reproduce the
$^3$H or $^3$He energy with several modern $\vij$.  More recently 
\cite{NKG00} they have obtained results for $^3$H, $^3$He and $^4$He 
with AV18 and a revised version of TM $\Vtpi$ as well as the UIX
model.  They observe that when the energy of $^3$He 
is reproduced that of $^4$He is very close to the observed, but $^3$H 
is underbound by $\sim 0.5\%$.  On the other hand if $^3$H energy is 
reproduced both $^3$He and $^4$He are overbound by less than 1\%. 
In order to improve the accuracy of nuclear Hamiltonians beyond 
99\%, a more quantitative description of the charge symmetry 
breaking interactions is necessary.  

The above results limit the contribution of four-nucleon interactions in 
$^4$He to less than 1\% of its binding energy.
The $\langle \Vijk \rangle$ contributes approximately a quarter of the $^4$He 
binding, although the difference of the total energies computed with and 
without a $\Vijk$ is only a seventh of the binding energy. 
The ratio of the expectation values of $\Vijk$ and $\vij$ is $\sim$ 7\% 
in the $A=8$ nuclei, and $\langle \Vijk \rangle$ gives up to one
half of their binding energy.  
If the four-nucleon interactions were to contribute 
$\sim$ 6\% of $\Vijk$, they could influence the $A=8$ binding energies 
by $\sim$ 3\%.  The present calculations do not indicate a need to 
include four-nucleon interactions to fit the observed energies 
at the 1\% level.  Thus, either the four-nucleon contributions are 
smaller than 1\% of the binding energies, 
or parts of the present models of $\Vijk$ 
are mocking up their effects. 

The energies of light nuclei can be used to determine at most three parameters
of $\Vijk$.  We can choose them as either the strengths $\Atpip$, $\Athpi$
and $\AR$, or use a theoretical value of $\Atpip$ and fit the short-range cutoff.
It is possible to make correlated changes in the three strengths, as in models IL2 and 
IL4, which have relatively small effect on the energies of light nuclei. 

All the realistic models of $\Vijk$, including the older Urbana models, have
a cancellation between the attractive $\Vtpi$ and repulsive $V^{R}$.
For example, their contributions in the AV18/IL2 model,
to the energy of $^3$He ($^8$Be)
are respectively $-2.8$ ($-37$) and 1.3 (18) MeV.  The contribution of $V^{R}$
grows faster than that of $\Vtpi$ as either $A$ or the density of matter increases,
and lowers the saturation density of symmetric nuclear matter.  
It is difficult to determine the $\AR$ quantitatively from the energies of 
light nuclei.  

Such a cancellation
was noticed in Faddeev calculations including $\Delta$ components
in the triton wave function~\cite{sauer,prb}.  The estimates of Picklesimer, Rice
and Brandenburg~\cite{prb} indicated that the $\Delta$ part of $\Vtpip$
changes the $^3$H energy by only $\sim -0.75$ MeV, 
while the processes, which represent
the suppression of the attractive two-nucleon $v^{2\pi}$ by the third
nucleon via the dispersion effect, give $\sim 1.1$ MeV.
The $V^{R}$ contribution in present models is comparable to their $\Delta$
dispersive effect, while the $\Vtpip$ is much more attractive. 

Studies of $A=3,4$ nuclei with relativistic Hamiltonians~\cite{rel2}
indicate that the
boost interaction $\dv$ gives the largest relativistic 
correction, of $\sim$ 0.4 MeV, to
the triton energy.  It is included in the present relativistic ($H^*$) models.  The other
corrections included in the relativistic Hamiltonians, but excluded
here, are only $\sim 0.1
\pm 0.05 $ MeV, {\em i.e.} of order 1\%.  However, three-nucleon
interactions via Z-diagrams~\cite{zone}, 
if any, have to be added to the $\Vijk ^*$
in the relativistic Hamiltonians.  Forest~\cite{junpc} 
has estimated their contribution
to the triton, using the scalar and vector parts of AV18 $\vij$, obtained
with Riska's method, to be $\sim 0.3$ MeV.  In the present Illinois models
these are also buried in the $V^{R*}$.  In the initial Illinois
models discussed at the
International Nuclear Physics Conference 
in Paris~\cite{VRPP99} we attempted to include possible Z-diagram
contributions in $\Vijk$, however,
the observed energies can be reproduced without them.  If these 
exist, then, as in the case of
$\Vtpis$, we can assume  theoretical values for their strengths and
fit the energies, presumably with $\sim$ 1\% accuracy, by readjusting
the strengths $\Atpip$, $\Athpi$ and $\AR^*$.  Nuclear binding energies  
seem to require only three components in the $\Vijk ^*$, an attractive 
part to provide more binding to light nuclei, a repulsive part to make 
nuclear matter $E(\rho)$ saturate at empirical $\rho_0$, and an 
isospin-dependent term to provide extra 
binding to the neutron-rich helium isotopes. 

Preliminary versions of the IL1-IL3 models were also discussed at the 
Few-Body Physics conference in Taipei~\cite{VRPP00}.  The parameters 
of the Illinois models reported there were incorrect due to a programming
error.  The correct values are as reported here.  Since then the IL3 
model has been revised.  
In the present IL3 model, softer cutoffs are used also
in the $\VR$ terms.  Results for nuclear and neutron matter with the 
Illinois models will be published separately. 

As mentioned in the introduction,
additional data necessary to further the study of $\Vijk$ may be 
obtained from the scattering of nucleons by deuterium. 
It is known that most of the low energy $Nd$ elastic scattering 
observables are well reproduced by realistic models of $\vij$~\cite{CS98}. 
The expectation value of $\Vijk$ is only $\sim 2.5\%$ of that of $\vij + 
v_{jk} + v_{ki} $ in $^3$H, thus it is not expected to have 
a large effect on this scattering.  However, all realistic models 
of $\vij$ underestimate the observed nucleon analyzing power $A_y$ in 
low energy $Nd$ scattering; the spin-orbit splitting induced by the 
$\Vtpip$ of the various $\Vijk$ reduces the error somewhat but
the additional spin-orbit splitting induced by the present Illinois models
is probably inadequate to completely correct the underestimate~\cite{pisak}.

More recently it has also been suggested that minima 
and polarization observables of $Nd$ 
elastic scattering at intermediate energies are 
sensitive to $\Vijk$, and may be used to refine models of $\Vijk$~\cite{ndscat,Sak00,Cad00}.
Thus further improvements in realistic models of $\Vijk$ may be possible by a 
simultaneous fit to $Nd$ scattering observables and nuclear binding energies.  
Since $Nd$ scattering is sensitive only to the $\Vijk$ in the total isospin 
$T = 1/2$ state, it does not provide information on the $\Vijk$ in the 
$T = 3/2$ state.  This work shows the need for an attractive interaction 
in $T=3/2$ states to reproduce the energies of $p$-shell nuclei. 
It may be possible to access this part of $\Vijk$ in $n-t$ and $p-^3$He 
scattering.

\acknowledgments

The many-body calculations were made possible by generous grants of time
on the parallel computers of the Mathematics and Computer Science
Division, Argonne National Laboratory, and by early-user time on the IBM
SP at the National Energy Research Scientific Computing Center.  The work 
of SCP and RBW is supported by the U.S. Department of Energy, Nuclear Physics
Division, under contract No. W-31-109-ENG-38, that of VRP by the U.S. 
National Science Foundation via grant PHY 98-00978, and that of JC by
the U.S. Department of Energy under contract W-7405-ENG-36.

\appendix

\section*{The Spin-Space Operators in $\Vthpi$ }

It is convenient to define the functions:
\begin{eqnarray}
t_{ij} &=& \frac{3 T(m_{\pi}r_{ij})}{r^2_{ij}}\ , \\ 
y_{ij} &=& Y(m_{\pi}r_{ij}) - T(m_{\pi}r_{ij}) \ ,
\end{eqnarray}
with which the spin-space operator $X_{ij}$ in $v^{\pi}_{ij}$ 
[Eq.(\ref{eq:vpi})] can be expressed as: 
\begin{equation}
X_{ij} = t_{ij} \boldsigma_i \cdot {\bf r}_{ij} ~ 
\boldsigma_j \cdot {\bf r}_{ij} + y_{ij} \boldsigma_i \cdot \boldsigma_j \ . 
\end{equation}
Evaluating the $\Delta$-ring diagrams using closure approximation then 
gives:
\begin{eqnarray}
S^I_{\sigma} &=& 2 y_{ij}y_{jk}y_{ki} + \case{2}{3} \sum_{cyc}
(r^2_{ij}t_{ij}y_{jk}y_{ki} + C_j^2 t_{ij}t_{jk}y_{ki})
-\case{2}{3} C_iC_jC_k t_{ij}t_{jk}t_{ki}  \nonumber \\ 
&+& \left[ \sum_{cyc} \boldsigma_i \cdot \boldsigma_j \right] 
\left[ \case{2}{3} y_{ij}y_{jk}y_{ki} + \case{1}{3} 
\sum_{cyc} r^2_{ij}t_{ij}y_{jk}y_{ki} \right] + \case{1}{3} 
\sum_{cyc} \boldsigma_i \cdot \boldsigma_k C_j^2 t_{ij}t_{jk}y_{ki} 
\nonumber \\
&-& \case{1}{3} \sum_{cyc} \left(\boldsigma_i \cdot {\bf r}_{ij}  ~
\boldsigma_j \cdot {\bf r}_{ij} ~ t_{ij}y_{ki}y_{jk}
                               + \boldsigma_i \cdot {\bf r}_{ki}  ~
\boldsigma_j \cdot {\bf r}_{ki} ~ t_{ki}y_{jk}y_{ij}
                               + \boldsigma_i \cdot {\bf r}_{jk}  ~
\boldsigma_j \cdot {\bf r}_{jk} ~ t_{jk}y_{ij}y_{ki}\right) \nonumber \\
&+& \case{1}{3} \sum_{cyc}  C_k \boldsigma_i \cdot {\bf r}_{jk} ~
\boldsigma_j \cdot {\bf r}_{ki} t_{ki}t_{jk}y_{ij} \nonumber \\
&+& \case{1}{3} \sum_{cyc} 
\boldsigma_i \cdot {\bf a} ~ \boldsigma_j \cdot {\bf a}
\left(t_{ij}t_{jk}y_{ki}+t_{ij}y_{jk}t_{ki} +C_k t_{ij}t_{jk}t_{ki} \right) 
\ ,
\end{eqnarray}
\begin{eqnarray}
S^D_{\sigma,ijk}&=& \case{1}{3}\boldsigma_j \cdot \boldsigma_k \left[ 
2 y_{ij}y_{jk}y_{ki} + C_i^2 t_{ij}y_{jk}t_{ki}
+ r^2_{ij}t_{ij}y_{ki}y_{jk} + r^2_{jk}t_{jk}y_{ij}y_{ki}
+ r^2_{ki}t_{ki}y_{jk}y_{ij} \right] \nonumber \\ 
&+& \case{1}{3}C_i ~ \boldsigma_j \cdot {\bf r}_{ki} ~ \boldsigma_k \cdot {\bf r}_{ij} ~ t_{ij}y_{jk}t_{ki}
 -  \case{1}{3} \boldsigma_j \cdot {\bf r}_{ij} ~ \boldsigma_k \cdot {\bf r}_{ij} ~ t_{ij}y_{ki}y_{jk} \nonumber  \\
&-& \case{1}{3} \boldsigma_j \cdot {\bf r}_{ki} ~ \boldsigma_k \cdot {\bf r}_{ki} ~ t_{ki}y_{jk}y_{ij} 
 -  \case{1}{3} \boldsigma_j \cdot {\bf r}_{jk} ~ \boldsigma_k \cdot {\bf r}_{jk} ~ t_{jk}y_{ij}y_{ki} \nonumber  \\
&+&\case{1}{3} \boldsigma_j \cdot {\bf a} ~ \boldsigma_k \cdot {\bf a}
\left(t_{ij}t_{jk}y_{ki}+y_{ij}t_{jk}t_{ki}+C_it_{ij}t_{jk}t_{ki} \right) \ ,
\end{eqnarray}
\begin{eqnarray}
A^I_{\sigma}&=&\case{i}{3} \left[ \boldsigma_i\cdot \boldsigma_j \times 
\boldsigma_k ~ y_{ij}y_{jk}y_{ki} + 
\boldsigma_i \cdot {\bf a} ~ \boldsigma_j \cdot {\bf a} ~ \boldsigma_k \cdot {\bf a} ~ 
t_{ij}t_{jk}t_{ki} \right] \nonumber  \\
&+&\case{i}{3} \sum_{cyc} \left( \boldsigma_i \times \boldsigma_j \cdot {\bf r}_{ij} ~ 
\boldsigma_k \cdot {\bf r}_{ij} ~ t_{ij}y_{jk}y_{ki}
+ \boldsigma_i \cdot {\bf a} ~ \boldsigma_j \cdot \boldsigma_k ~ C_i 
t_{ij}y_{jk}t_{ki} \right) \nonumber \\
&+&\case{i}{3}\sum_{cyc} \boldsigma_i \cdot {\bf r}_{jk} ~ \boldsigma_k \cdot {\bf r}_{ij} ~
\boldsigma_j \cdot {\bf a} ~ t_{ij}t_{jk}y_{ki}  \nonumber \\
&+& \case{2i}{3} \sum_{cyc} \boldsigma_i \cdot {\bf a}\left(C_it_{ij}y_{jk}t_{ki} 
- C_jt_{ij}t_{jk}y_{ki} -C_k y_{ij}t_{jk}t_{ki} - C_jC_kt_{ij}t_{jk}t_{ki}
\right)  \ ,
\label{eq:AIsig}
\end{eqnarray}
and 
\begin{equation}
A^D_{\sigma,ijk}=-\case{i}{3}\boldsigma_i \cdot {\bf a}\left(C_it_{ij}y_{jk}t_{ki} 
- C_jt_{ij}t_{jk}y_{ki} -C_k y_{ij}t_{jk}t_{ki} - C_jC_kt_{ij}t_{jk}t_{ki}
\right) \ . 
\end{equation}
The $C_i = {\bf r}_{ij} \cdot {\bf r}_{ik} $
and ${\bf a}={\bf r}_{ij} \times {\bf r}_{jk}$.  
A cyclic sum over the indices $ijk$ is denoted by $cyc$.

\begin{table}
\caption{Three-body potential parameters used in this paper.  Parameters that were not
varied in fitting the data are marked with an $^*$.}
\begin{tabular}{lddddddd}
Model & $c $    & $  \Atpip$    & $\Atpis$   & $\Athpi$    & $\AR$       & $A_{W}$  &  $\AR^*$  \\
    & fm$^{-2}$ &    MeV      &  MeV       &  MeV        &  MeV        &  MeV     &   MeV  \\
\hline                                                                   
UIX   & 2.1$^*$ &  $-$0.0293    &   $-$      &  $-$        & 0.00480     &   0.$^*$ & 0.00291 \\
IL1   & 2.1$^*$ &  $-$0.0385    &  0.0$^*$   &  0.0026$^*$ & 0.00705$^*$ &   0.$^*$ & 0.00491 \\
IL2   & 2.1$^*$ &  $-$0.037     & $-$1.0$^*$ &  0.0026     & 0.00705     &   0.$^*$ & 0.00493 \\
IL3   & 1.5     &  $-$0.07$^*$  & $-$1.0$^*$ &  0.0065     & 0.032       &   0.$^*$ & 0.02562 \\
IL4   & 2.1$^*$ &  $-$0.028$^*$ & $-$1.0$^*$ &  0.0021     & 0.0039      &   0.$^*$ & 0.00196 \\
IL5   & 2.1$^*$ &  $-$0.03      & $-$1.0$^*$ &  0.0021$^*$ & 0.002$^*$   & 210.     & 0.0     \\
\end{tabular}
\label{table:params}
\end{table}
         
\begin{table}
\squeezetable
\caption{Experimental and GFMC energies (in MeV) of particle-stable or 
narrow-width nuclear states and of neutron drops.  Monte Carlo statistical
errors in the last digits are shown in parentheses. The final column
gives experimental widths in keV.}

\begin{tabular}{lccccccccrr}
                        &   $\avep$    & $\avet$      &       UIX    &       IL1    &       IL2    &       IL3    &      IL4     &      IL5     & Expt. & $\Gamma$~\\
\hline 
$^3$H($\case{1}{2}^+$)  & ~$-$7.76(1)  & ~$-$7.61(1)  & ~$-$8.46(1)~ & ~$-$8.43(1)~ & ~$-$8.43(1)  & ~$-$8.41(1)  & ~$-$8.44(1)  & ~$-$8.41(1)  & ~$-$8.48  &      \\
$^3$He($\case{1}{2}^+$) & ~$-$7.02(1)  & ~$-$6.87(1)  & ~$-$7.71(1)~ & ~$-$7.68(1)  & ~$-$7.67(1)  & ~$-$7.66(1)  & ~$-$7.69(1)  & ~$-$7.66(1)  & ~$-$7.72  &      \\
$^4$He(0$^+$)           & $-$25.14(2)~ & $-$24.07(4)~ & $-$28.33(2)~ & $-$28.38(2)~ & $-$28.37(3)~ & $-$28.24(3)~ & $-$28.35(2)~ & $-$28.23(2)~ & $-$28.30 &      \\
$^6$He(0$^+$)           & $-$25.20(6)~ & $-$23.9(1)~~ & $-$28.1(1)~~ & $-$29.4(1)~~ & $-$29.4(1)~~ & $-$29.3(2)~~ & $-$29.3(1)~~ & $-$29.5(1)~~ & $-$29.27 &      \\
$^6$He(2$^+$)           & $-$23.18(6)~ & $-$21.8(1)~~ & $-$26.3(1)~~ & $-$27.2(1)~~ & $-$27.1(1)~~ & $-$27.8(1)~~ & $-$27.4(1)~~ & $-$27.3(1)~~ & $-$27.47 &  113 \\
$^6$Li(1$^+$)           & $-$28.19(5)~ & $-$26.9(1)~~ & $-$31.1(1)~~ & $-$31.9(1)~~ & $-$32.3(1)~~ & $-$32.2(1)~~ & $-$32.0(1)~~ & $-$32.1(1)~~ & $-$31.99 &      \\
$^6$Li(3$^+$)           & $-$24.98(5)~ & $-$23.5(1)~~ & $-$28.1(1)~~ & $-$30.1(2)~~ & $-$30.1(2)~~ & $-$30.0(2)~~ & $-$29.8(2)~~ & $-$29.8(2)~~ & $-$29.80 &   24 \\
$^7$He($\case{3}{2}^-$) & $-$22.82(10) & $-$21.2(2)~~ & $-$25.8(2)~~ & $-$29.3(3)~~ & $-$29.2(3)~~ & $-$29.3(3)~~ & $-$29.3(3)~~ & $-$29.2(2)~~ & $-$28.82 &  160 \\
$^7$Li($\case{3}{2}^-$) & $-$33.56(6)~ & $-$31.6(1)~~ & $-$37.8(1)~~ & $-$39.4(2)~~ & $-$39.6(2)~~ & $-$39.3(2)~~ & $-$39.5(2)~~ & $-$39.3(2)~~ & $-$39.24 &      \\
$^7$Li($\case{1}{2}^-$) & $-$33.17(7)~ & $-$31.1(2)~~ & $-$37.5(2)~~ & $-$39.2(2)~~ & $-$39.1(2)~~ & $-$38.7(2)~~ & $-$39.0(2)~~ & $-$39.0(2)~~ & $-$38.77 &      \\
$^7$Li($\case{7}{2}^-$) & $-$28.41(6)~ & $-$26.4(1)~~ & $-$32.1(1)~~ & $-$34.5(3)~~ & $-$34.4(3)~~ & $-$34.0(2)~~ & $-$34.5(2)~~ & $-$34.2(3)~~ & $-$34.61 &   93 \\
$^8$He(0$^+$)           & $-$23.8(1)~~ & $-$21.6(2)~~ & $-$27.2(2)~~ & $-$30.5(3)~~ & $-$31.3(3)~~ & $-$32.0(4)~~ & $-$31.9(4)~~ & $-$31.0(2)~~ & $-$31.41 &      \\
$^8$Li(2$^+$)           & $-$34.2(1)~~ & $-$31.8(3)~~ & $-$38.0(2)~~ & $-$41.8(3)~~ & $-$42.2(2)~~ & $-$41.2(3)~~ & $-$42.0(3)~~ & $-$42.5(3)~~ & $-$41.28 &      \\
$^8$Li(1$^+$)           & $-$33.9(1)~~ & $-$31.6(2)~~ & $-$37.4(2)~~ & $-$40.5(3)~~ & $-$40.5(3)~~ & $-$40.2(3)~~ & $-$40.9(3)~~ & $-$40.9(3)~~ & $-$40.30 &      \\
$^8$Li(3$^+$)           & $-$31.4(1)~~ & $-$28.9(2)~~ & $-$35.3(2)~~ & $-$39.3(3)~~ & $-$39.1(3)~~ & $-$39.5(4)~~ & $-$39.3(3)~~ & $-$39.2(3)~~ & $-$39.02 &   33 \\
$^8$Li(4$^+$)           & $-$28.1(1)~~ & $-$25.5(2)~~ & $-$31.7(2)~~ & $-$34.9(3)~~ & $-$35.0(3)~~ & $-$34.7(3)~~ & $-$35.2(3)~~ & $-$34.9(3)~~ & $-$34.75 &   35 \\
$^8$Be(0$^+$)           & $-$47.9(1)~~ & $-$45.6(3)~~ & $-$54.4(2)~~ & $-$57.2(4)~~ & $-$56.6(4)~~ & $-$55.6(4)~~ & $-$56.5(3)~~ & $-$55.7(3)~~ & $-$56.50 &      \\
$^8$Be(1$^+$)           & $-$32.8(2)~~ & $-$30.9(3)~~ & $-$36.3(3)~~ & $-$37.8(2)~~ & $-$37.6(2)~~ & $-$37.3(3)~~ & $-$38.8(3)~~ & $-$38.9(4)~~ & $-$38.35 &  138 \\
\hline
$^7$n($\case{1}{2}^-$)  & $-$33.78(4)  & $-$33.47(5)  & $-$33.2(1)~  & $-$36.0(2)~   & $-$35.8(2)~   & $-$36.6(3)~   & $-$35.2(3)~   & $-$35.3(3)~   &   &       \\
$^7$n($\case{3}{2}^-$)  & $-$32.25(4)  & $-$31.82(5)  & $-$31.7(1)~  & $-$33.2(2)~   & $-$33.0(2)~   & $-$33.0(3)~   & $-$32.9(3)~   & $-$33.1(2)~   &   &       \\
$^8$n(0$^+$)            & $-$39.73(6)  & $-$39.21(8)  & $-$37.8(1)~  & $-$41.3(3)~   & $-$41.1(3)~   & $-$40.7(2)~   & $-$40.7(3)~   & $-$40.7(2)~   &   &       \\
\end{tabular}

\label{table:narrow-energies}
\end{table}

\begin{table}
\caption{Average deviations (in MeV) from experimental energies.  For each 
Hamiltonian, the average signed deviation, average magnitude of deviation and
rms deviation are shown for the 17 ``narrow'' states given
in Table~\protect\ref{table:narrow-energies} (Only $^3$He energies are used
for $A=3$).}

\begin{tabular}{lrddd}
Model    &   Average   &     Average     &    rms     \\
         &  Deviation  & $|$Deviation$|$ & Deviation  \\
\hline 
$\avep$  &     5.52(2) &     5.52       &   5.83 \\
$\avet$  &     7.32(5) &     7.32       &   7.72 \\
AV18/UIX &     2.02(4) &     2.02       &   2.34 \\
AV18/IL1 &  $-$0.09(6) &     0.31       &   0.38 \\
AV18/IL2 &  $-$0.10(6) &     0.28       &   0.36 \\
AV18/IL3 &     0.04(7) &     0.31       &   0.44 \\
AV18/IL4 &  $-$0.21(6) &     0.24       &   0.33 \\
AV18/IL5 &  $-$0.12(6) &     0.34       &   0.46 \\
\end{tabular}
\label{table:energy-errors}
\end{table}

\begin{table}
\caption{Average deviations (in MeV) from experimental excitation energies
for the 8 ``narrow'' excited states.  
As in Table~\protect\ref{table:energy-errors}, but for excitation
energies rather than total energies.}

\begin{tabular}{lrdddd}
Model    &   Average   &     Average     &    rms    \\
         &  Deviation  & $|$Deviation$|$ & Deviation \\
\hline 
$\avep$  &  $-$0.23(5)  &     0.83       &   1.20 \\
$\avet$  &  $-$0.22(10) &     0.90       &   1.36 \\
AV18/UIX &     0.17(8)  &     0.41       &   0.53 \\
AV18/IL1 &     0.29(13) &     0.44       &   0.53 \\
AV18/IL2 &     0.53(12) &     0.53       &   0.61 \\
AV18/IL3 &     0.03(14) &     0.24       &   0.34 \\
AV18/IL4 &     0.09(12) &     0.20       &   0.25 \\
AV18/IL5 &     0.27(13) &     0.66       &   0.79 \\
\end{tabular}
\label{table:excitation-errors}
\end{table}
         
\begin{table}
\caption{Ground-state expectation values of the two-body potential and $\langle K+\vij \rangle$
(in MeV) for the \avep, \avet, AV18/UIX, and AV18/IL2 Hamiltonians.}
\begin{tabular}{lddddddd}
       & $\avep$    & \multicolumn{2}{c}{$\avet$}& \multicolumn{2}{c}{AV18/UIX}& \multicolumn{2}{c}{AV18/IL2} \\
       & $v_{ij}$~~  & $v_{ij}$~~  & $K+v_{ij}$   & $v_{ij}$~~  & $K+v_{ij}$  & $v_{ij}$~~  & $K+v_{ij}$ \\
\hline
$^3$H  &  $-$54.9(2) &  $-$54.8(2) &  $-$7.605(5) &  $-$58.7(2) &  $-$7.27(1) &  $-$58.6(2) &  $-$6.97(1) \\
$^3$He &  $-$53.4(2) &  $-$53.3(2) &  $-$6.882(5) &  $-$57.1(2) &  $-$6.54(1) &  $-$56.7(2) &  $-$6.27(1) \\
$^4$He & $-$126.(1)  & $-$124.9(7) & $-$24.07(4)  & $-$135.9(5) & $-$21.98(6) & $-$136.4(5) & $-$19.99(8) \\
$^6$He & $-$155.(1)  & $-$153.(1)  & $-$23.89(8)  & $-$164.(1)  & $-$21.1(2)  & $-$171.(2)  & $-$17.9(3)  \\
$^6$Li & $-$174.(1)  & $-$173.(1)  & $-$26.9(1)   & $-$182.(1)  & $-$23.8(2)  & $-$187.(2)  & $-$21.2(3)  \\
$^7$Li & $-$221.(2)  & $-$219.(2)  & $-$31.6(1)   & $-$225.(2)  & $-$28.7(2)  & $-$232.(3)  & $-$25.1(5)  \\
$^8$He & $-$193.(3)  & $-$191.(3)  & $-$21.6(2)   & $-$194.(1)  & $-$19.1(2)  & $-$218.(3)  & $-$15.(1)   \\
$^8$Li & $-$249.(4)  & $-$247.(3)  & $-$31.8(3)   & $-$255.(2)  & $-$27.8(3)  & $-$278.(2)  & $-$21.6(4)  \\
$^8$Be & $-$287.(3)  & $-$284.(3)  & $-$45.6(3)   & $-$297.(2)  & $-$39.6(4)  & $-$303.(3)  & $-$35.5(8)  \\
\hline
$^7$n  &  $-$54.8(7) &  $-$54.5(7) & $-$33.47(5)  &  $-$54.2(8) & $-$33.86(9) &  $-$59.(1)  & $-$32.2(4)  \\
$^8$n  &  $-$69.8(6) &  $-$69.3(6) & $-$39.21(8)  &  $-$65.9(8) & $-$38.8(1)  &  $-$73.(1)  & $-$38.2(5)  \\
\end{tabular}
\label{table:vij}
\end{table}

\begin{table}
\caption{Total three-nucleon potential energies (in MeV) for the
AV18/UIX and Illinois Hamiltonians.}
\begin{tabular}{ldddddd}
       &  AV18/UIX    &  AV18/IL1   &  AV18/IL2   &  AV18/IL3   &  AV18/IL4   &  AV18/IL5    \\
\hline
$^3$H  &  $-$1.19(1)  &  $-$1.46(1) &  $-$1.46(1) &  $-$1.65(1) &  $-$1.25(1) &  $-$1.24(1) \\
$^3$He &  $-$1.17(1)  &  $-$1.44(1) &  $-$1.41(1) &  $-$1.64(1) &  $-$1.22(1) &  $-$1.24(1) \\
$^4$He &  $-$6.35(5)  &  $-$8.4(1)  &  $-$8.38(7) & $-$10.02(7) &  $-$7.18(6) &  $-$7.24(5) \\
$^6$He &  $-$7.0(1)   & $-$11.3(2)  & $-$11.5(3)  & $-$14.1(3)  &  $-$9.8(2)  &  $-$9.9(2)  \\
$^6$Li &  $-$7.3(2)   & $-$11.2(3)  & $-$11.1(3)  & $-$13.6(3)  &  $-$9.6(2)  & $-$10.2(2)  \\
$^7$Li &  $-$9.1(2)   & $-$13.8(4)  & $-$14.5(4)  & $-$16.9(4)  & $-$12.8(4)  & $-$13.4(4)  \\
$^8$He &  $-$8.0(2)   & $-$15.5(5)  & $-$16.3(5)  & $-$19.7(6)  & $-$16.(1)   & $-$14.7(3)  \\
$^8$Li & $-$10.2(2)   & $-$19.3(5)  & $-$20.6(4)  & $-$24.9(6)  & $-$18.2(4)  & $-$17.3(5)  \\
$^8$Be & $-$14.9(3)   & $-$20.8(7)  & $-$21.(1)   & $-$25.1(8)  & $-$19.0(4)  & $-$19.4(5)  \\
\hline
$^7$n  &     0.69(4)  &  $-$3.8(3)  &  $-$3.6(3)  &  $-$4.8(4)  &  $-$2.5(3)  &  $-$2.5(4)  \\
$^8$n  &     1.01(6)  &  $-$3.1(4)  &  $-$3.0(4)  &  $-$3.7(3)  &  $-$2.8(3)  &  $-$1.9(2)  \\
\end{tabular}
\label{table:Vijk}
\end{table}

\begin{table}
\caption{Contributions of various three-body potential terms (in MeV)
evaluated for the AV18/IL2 Hamiltonian.}
\begin{tabular}{ldddddd}
       & $V^{2\pi,A}$& $V^{2\pi,C}$&  $\Vtpis$   & $V^{3\pi,SS}$ & $V^{3\pi,AA}$& $\VR$ \\
\hline
$^3$H  &  $-$1.78(1) &  $-$1.082(9)& $-$0.119(1) &       0.      & 0.182(3)   &   1.34(1) \\
$^3$He &  $-$1.72(1) &  $-$1.045(8)& $-$0.115(1) &       0.      & 0.176(3)   &   1.29(1) \\
$^4$He &  $-$9.76(8) &  $-$5.85(5) & $-$0.652(5) &       0.      & 0.63(1)    &   7.26(7) \\
$^6$He & $-$12.2(3)  &  $-$7.3(1)  & $-$0.74(2)  &    $-$1.33(6) & 0.42(4)    &   9.6(3)  \\
$^6$Li & $-$11.9(2)  &  $-$7.2(1)  & $-$0.72(2)  &    $-$0.81(4) & 0.37(4)    &   9.1(2)  \\
$^7$Li & $-$15.4(4)  &  $-$9.3(2)  & $-$0.91(3)  &    $-$1.61(9) & 0.5(1)     &  12.3(4)  \\
$^8$He & $-$15.6(4)  &  $-$9.1(2)  & $-$0.88(3)  &    $-$4.5(2)  & 0.44(7)    &  13.3(4)  \\
$^8$Li & $-$20.7(3)  & $-$12.2(2)  & $-$1.26(2)  &    $-$4.4(1)  & 0.55(5)    &  17.4(3)  \\
$^8$Be & $-$23.1(6)  & $-$14.0(3)  & $-$1.38(4)  &    $-$1.6(1)  & 0.7(1)     &  18.3(6)  \\
\hline                                                                         
$^7$n  &  $-$0.15(5) &             &    0.08(1)  &    $-$5.4(3)  & 0.         &   1.9(1)  \\
$^8$n  &     0.13(9) &             &    0.18(1)  &    $-$5.9(4)  & 0.         &  2.6(2)  \\
\end{tabular}
\label{table:IL2-terms}
\end{table}

\begin{table}
\caption{Contributions of various three-body potential terms (in MeV)
evaluated for the AV18/IL3 Hamiltonian.}
\begin{tabular}{ldddddd}
       & $V^{2\pi,A}$& $V^{2\pi,C}$&  $\Vtpis$   & $V^{3\pi,SS}$ & $V^{3\pi,AA}$& $\VR$ \\
\hline
$^3$H  &  $-$2.37(2) &  $-$1.47(1) &  $-$0.092(1) &    0.      & 0.217(3)   &    2.07(2) \\
$^3$He &  $-$2.35(2) &  $-$1.46(1) &  $-$0.091(1) &    0.      & 0.212(3)   &    2.04(2) \\
$^4$He & $-$12.99(9) &  $-$7.95(5) &  $-$0.525(4) &    0.      & 0.54(1)    &   10.9(1)  \\
$^6$He & $-$16.4(3)  &  $-$9.9(2)  &  $-$0.61(1)  & $-$1.9(1)  & 0.26(4)    &   14.5(3)  \\
$^6$Li & $-$16.2(3)  &  $-$9.9(2)  &  $-$0.59(1)  & $-$1.31(6) & 0.25(5)    &   14.2(4)  \\
$^7$Li & $-$20.9(5)  & $-$12.8(3)  &  $-$0.73(2)  & $-$2.32(9) & 0.38(7)    &   19.5(6)  \\
$^8$He & $-$21.6(5)  & $-$12.6(3)  &  $-$0.78(3)  & $-$7.0(2)  & 0.16(9)    &   22.2(7)  \\
$^8$Li & $-$27.8(6)  & $-$16.4(3)  &  $-$0.96(2)  & $-$6.2(2)  & 0.13(9)    &   26.3(7)  \\
$^8$Be & $-$30.8(7)  & $-$18.9(4)  &  $-$1.12(3)  & $-$2.4(2)  & 0.7(1)     &   27.4(9)  \\
\hline                                                                        
$^7$n  &  $-$0.13(8) &             &     0.07(1)  & $-$8.5(5)  & 0.         &    3.8(2)  \\
$^8$n  &     0.52(8) &             &     0.17(1)  & $-$8.9(4)  & 0.         &    4.5(1)  \\
\end{tabular}
\label{table:IL3-terms}
\end{table}

\begin{table}
\caption{Contributions of two-nucleon and three-nucleon pion and remainder
potentials (in MeV) for the AV18/IL2 Hamiltonian.}
\begin{tabular}{lddddd}
       &  $\vpi$      &  $\vr$     &  $\Vtpi$    &  $V^{3\pi}$ &  $\VR$  \\
\hline
$^3$H  &  $-$45.0(2)  & $-$13.5(2) &  $-$2.98(2) &    0.182(3) &  1.34(1) \\
$^3$He &  $-$44.4(2)  & $-$12.4(2) &  $-$2.88(2) &    0.176(3) &  1.29(1) \\
$^4$He & $-$105.4(4)  & $-$30.9(5) & $-$16.3(1)  &    0.63(1)  &  7.26(7) \\
$^6$He & $-$127.(1)   & $-$44.(2)  & $-$20.3(4)  & $-$0.91(6)  &  9.6(3)  \\
$^6$Li & $-$150.(1)   & $-$38.(2)  & $-$19.8(4)  & $-$0.44(5)  &  9.1(2)  \\
$^7$Li & $-$178.(2)   & $-$54.(3)  & $-$25.6(6)  & $-$1.1(1)   & 12.3(4)  \\
$^8$He & $-$153.(1)   & $-$66.(3)  & $-$25.6(6)  & $-$4.0(2)   & 13.3(4)  \\
$^8$Li & $-$211.(1)   & $-$67.(2)  & $-$34.2(5)  & $-$3.8(1)   & 17.4(3)  \\
$^8$Be & $-$234.(2)   & $-$69.(3)  & $-$38.5(9)  & $-$0.9(2)   & 18.3(6)  \\
\hline
$^7$n  &  $-$10.11(9) & $-$49.(1)  &  $-$0.07(5) & $-$5.4(3)   &  1.9(1)  \\
$^8$n  &  $-$12.0(1)  & $-$61.(1)  &     0.31(9) & $-$5.9(4)   &  2.6(2)  \\
\end{tabular}
\label{table:IL2-vpi-r}
\end{table}

\begin{table}
\caption{Isovector and isotensor energies ($a^{(n)}_{A,T}$ in keV) for isomultiplets.}
\begin{tabular}{cccrrrrrrrr}
$A$&$T$           &$n$& $\avet$~& AV18/UIX  & AV18/IL1  & AV18/IL2  & AV18/IL3  & AV18/IL4  & AV18/IL5  &  Expt.\\
\hline                                                      
3 & $\case{1}{2}$ & 1 &  732(1) &  762(1)~~ &  757(2)~~ &  757(1)~~ &  752(1)~~ &  760(2)~~ &  759(2)~~ &   764 \\
3 & $\case{1}{2}$ & 1 &  723(2) &  753(1)~~ &  748(2)~~ &  747(1)~~ &  746(1)~~ &  751(2)~~ &  750(2)~~ &   764 \\
3 & $\case{1}{2}$ & 1 &  731(7) &  753(8)~~ &  754(7)~~ &  763(7)~~ &  748(8)~~ &  754(7)~~ &  753(7)~~ &   764 \\
\hline                                  
6 & $1$           & 1 & 1068(4) & 1102(9)~~ & 1149(6)~~ & 1172(6)~~ & 1141(5)~~ & 1147(6)~~ & 1179(6)~~ &  1173 \\
7 & $\case{1}{2}$ & 1 & 1586(7) & 1565(7)~~ & 1613(8)~~ & 1588(7)~~ & 1554(9)~~ & 1609(8)~~ & 1610(8)~~ &  1644 \\
8 & $2$           & 1 & 1457(7) & 1488(5)~~ & 1581(7)~~ & 1622(8)~~ & 1631(9)~~ & 1670(9)~~ & 1637(6)~~ &  1659 \\
8 & $1$           & 1 & 1636(9) & 1672(7)~~ & 1762(8)~~ & 1810(6)~~ & 1759(8)~~ & 1837(7)~~ & 1774(9)~~ &  1770 \\
\hline                                                                         
6 & $1$           & 2 &         & 251(18)~  & 275(13)   & 293(13)   & 266(16)   & 287(18)   & 280(17)   &  223 \\
8 & $2$           & 2 &         & 158(4)~~  & 170(4)~~  & 180(5)~~  & 180(5)~~  & 188(5)~~  & 175(3)~~  &  153 \\
8 & $1$           & 2 &         & 135(7)~~  & 135(12)   & 143(8)~~  &  99(12)   & 148(10)   & 156(12)   &  145 \\
\end{tabular}

\label{table:isoenergies}
\end{table}

\begin{table}
\caption{Various contributions to the isovector and isotensor energies
(in keV) computed with AV18/IL2. The definitions of the contributions are given
in the text.}

\begin{tabular}{cccrrrdrr}
$A$&$T$           &$n$& $K^{CSB}$& $v_{C1}(pp)$ & $v^{\gamma,R}$
                                                        & $v^{CSB}+v^{CD}$ & Total~~~ & Expt.\\
\hline
3 & $\case{1}{2}$ & 1 &   14(0)  &     649(1)   &    29(0)   &    64(0)~ &  757(1)~ &   764\\ 
6 & $1$           & 1 &   16(0)  &    1091(5)   &    18(0)   &    47(1)~ & 1172(6)~ &  1173 \\
7 & $\case{1}{2}$ & 1 &   22(0)  &    1447(6)   &    40(0)   &    79(2)~ & 1588(7)~ &  1644 \\
8 & $2$           & 1 &   18(0)  &    1528(7)   &    17(0)   &    59(1)~ & 1622(8)~ &  1659 \\
8 & $1$           & 1 &   23(0)  &    1686(5)   &    24(0)   &    76(1)~ & 1810(6)~ &  1770 \\
\hline
6 & $1$           & 2 &          &     166(1)   &    19(0)   &   107(13) & 293(13)  &  223 \\
8 & $2$           & 2 &          &     136(1)   &     6(0)   &    38(5)~ & 180(5)~  &  153 \\
8 & $1$           & 2 &          &     141(1)   &     4(0)   &  $-$3(8)~ & 143(8)~  &  145 \\
\end{tabular}
\label{table:IL2-isoenergies}
\end{table}

\begin{table}
\squeezetable
\caption{Experimental and computed energies (in MeV) of ``wide'' or experimentally 
unknown states.  The experimental widths (in keV) of the states are also
given.}
\begin{tabular}{lcccccccclr}
                        &   $\avep$    & $\avet$     &      UIX   &     IL1    &     IL2    &     IL3    &    IL4     &      IL5   & ~~Expt.  & $\Gamma$~\\
\hline                                                                                                                            
$^5$He($\case{3}{2}^-$) & $-$23.85(4)~ & $-$22.47(9) & $-$26.9(1) & $-$27.7(1) & $-$27.7(1) & $-$27.4(1) & $-$27.5(1) & $-$27.4(1) & $-$27.52(2)~ &  650 \\
$^5$He($\case{1}{2}^-$) & $-$23.17(3)~ & $-$21.9(1)~ & $-$25.8(1) & $-$26.5(1) & $-$26.4(1) & $-$26.3(1) & $-$26.1(1) & $-$26.0(1) & $-$26.32(20) & 5500 \\
$^6$He(1$^+$)           & $-$21.58(4)~ & $-$20.2(1)~ & $-$24.4(1) & $-$24.7(1) & $-$24.5(1) & $-$24.2(1) & $-$24.1(2) & $-$24.1(1) &              &      \\
$^6$Li(2$^+$)           & $-$24.12(4)~ & $-$22.7(1)~ & $-$27.2(1) & $-$27.9(1) & $-$27.9(1) & $-$27.7(2) & $-$27.9(1) & $-$27.8(1) & $-$27.68(2)~ & 1700 \\
$^7$He($\case{1}{2}^-$) & $-$22.01(10) & $-$20.8(2)~ & $-$24.3(2) & $-$26.6(2) & $-$26.5(2) & $-$26.3(2) & $-$26.1(2) & $-$26.3(2) &              &      \\
$^7$He($\case{5}{2}^-$) & $-$20.81(10) & $-$19.2(2)~ & $-$23.2(2) & $-$24.7(3) & $-$24.4(3) & $-$25.0(2) & $-$25.0(2) & $-$25.0(2) & $-$25.92(30) & 2200 \\
$^7$Li($\case{5}{2}^-$) & $-$27.52(5)~ & $-$25.7(1)~ & $-$31.3(1) & $-$32.3(2) & $-$32.2(2) & $-$32.0(2) & $-$32.1(2) & $-$32.3(2) & $-$32.56(5)~ &  875 \\
$^8$He(2$^+$)           & $-$21.39(8)~ & $-$19.6(2)~ & $-$24.1(2) & $-$26.8(3) & $-$26.6(3) & $-$26.2(3) & $-$27.2(3) & $-$26.6(3) & $-$27.82(5)~ &  630 \\
$^8$He(1$^+$)           & $-$21.2(1)~~ & $-$19.6(2)~ & $-$22.7(2) & $-$26.0(3) & $-$25.8(3) & $-$26.2(3) & $-$25.8(3) & $-$25.8(3) &              &      \\
$^8$Li(0$^+$)           & $-$33.5(1)~~ & $-$31.3(2)~ & $-$36.1(2) & $-$38.4(3) & $-$38.4(3) & $-$37.2(4) & $-$37.8(4) & $-$38.3(4) &              &      \\
$^8$Be(2$^+$)           & $-$45.6(1)~~ & $-$42.7(3)~ & $-$51.5(2) & $-$53.6(3) & $-$53.5(3) & $-$52.4(3) & $-$53.1(3) & $-$53.2(3) & $-$53.46(3)~ & 1500 \\
$^8$Be(4$^+$)           & $-$38.7(1)~~ & $-$36.2(2)~ & $-$44.9(2) & $-$45.5(3) & $-$45.4(3) & $-$45.0(3) & $-$45.4(3) & $-$45.9(3) & $-$45.10(30) & 3500 \\
$^8$Be(3$^+$)           & $-$31.2(2)~~ & $-$29.3(3)~ & $-$34.9(3) & $-$37.2(3) & $-$37.1(3) & $-$36.9(3) & $-$38.0(4) & $-$37.2(3) & $-$37.26(3)~ &  230 \\

\end{tabular}
\label{table:other-energies}
\end{table}

\begin{table}
\caption{Computed and experimental spin-orbit splittings in MeV.}
\begin{tabular}{lccccccccccccc}
       &                                    &$L$&  $S$          & \avep   & \avet  & UIX    & IL1    & IL2    & IL3    & IL4    & IL5    & Expt\\
\hline                                                         

$^5$He &  $\case{1}{2}^- - \case{3}{2}^-$ & 1 & $\case{1}{2}$ & 0.68(5)  & 0.6(1)~ & 1.1(2) & 1.2(2) & 1.3(2) & 1.1(2) & 1.4(2) & 1.3(2) & 1.20 \\
$^6$Li &  2$^+ - 3^+$                     & 2 &     1         & 0.86(6)  & 0.8(1)~ & 0.9(1) & 2.2(2) & 2.2(2) & 2.4(2) & 1.9(2) &        & 2.12 \\
$^7$Li &  $\case{1}{2}^- - \case{3}{2}^-$ & 1 & $\case{1}{2}$ & 0.39(9)  & 0.5(2)~ & 0.3(2) & 0.2(3) & 0.6(3) & 0.6(3) & 0.4(3) & 0.3(3) & 0.47 \\
$^7$Li &  $\case{5}{2}^- - \case{7}{2}^-$ & 3 & $\case{1}{2}$ & 0.89(8)  & 0.7(2)~ & 0.8(2) & 2.2(3) & 2.2(3) & 2.0(3) & 2.4(3) & 2.0(3) & 2.05 \\
$^8$Li &  1$^+ - 2^+$                     & 1 &     1         & 0.3(2)~  & 0.2(4)~ & 0.6(2) & 1.3(4) & 1.7(4) & 1.1(5) & 1.1(4) & 1.6(4) & 0.98 \\
\hline                                                                                     
$^7$n  &  $\case{3}{2}^- - \case{1}{2}^-$ & 1 & $\case{1}{2}$ & 1.53(6)  & 1.65(7) & 1.5(1) & 2.8(3) & 2.8(3) & 3.6(4) & 2.4(4) & 2.3(3) &      \\
\end{tabular}

\label{table:so-splits}
\end{table}

\begin{table}
\squeezetable
\caption{RMS point proton and neutron radii in fm.}
\begin{tabular}{lccccccccccc}
   & \multicolumn{2}{c}{$\avep$}&
                         \multicolumn{2}{c}{AV18/UIX}&
                                              \multicolumn{2}{c}{AV18/IL2}&
                                                                \multicolumn{2}{c}{AV18/IL3}& 
                                                                               \multicolumn{2}{c}{AV18/IL4} & Expt \\
       &  $p$    &  $n$    &  $p$    &  $n$    &  $p$    &  $n$    &  $p$    &  $n$    &  $p$    &  $n$     &  $p$ \\
\hline                                                                                                     
                                                                                                           
$^3$H  & 1.66(0) & 1.82(0) & 1.59(0) & 1.73(0) & 1.59(0) & 1.74(0) & 1.60(0) & 1.74(0) & 1.59(0) & 1.73(0)  & 1.60 \\
$^3$He & 1.85(0) & 1.68(0) & 1.76(0) & 1.61(0) & 1.76(0) & 1.61(0) & 1.76(0) & 1.61(0) & 1.76(0) & 1.61(0)  & 1.77 \\
$^4$He & 1.50(0) & 1.50(0) & 1.44(0) & 1.44(0) & 1.45(0) & 1.45(0) & 1.46(0) & 1.46(0) & 1.44(0) & 1.44(0)  & 1.47 \\
$^6$He & 2.06(1) & 3.07(1) & 1.97(1) & 2.94(1) & 1.91(1) & 2.82(1) & 1.99(1) & 2.97(1) & 1.99(1) & 2.96(1)  \\
$^6$Li & 2.50(1) & 2.50(1) & 2.57(1) & 2.57(1) & 2.39(1) & 2.39(1) & 2.44(1) & 2.44(1) & 2.38(1) & 2.38(1)  & 2.43 \\
$^7$Li & 2.29(1) & 2.47(1) & 2.33(1) & 2.52(1) & 2.25(1) & 2.44(1) & 2.32(1) & 2.52(1) & 2.26(1) & 2.44(1)  & 2.27 \\
$^8$He & 1.93(1) & 3.22(2) & 1.98(0) & 3.17(1) & 1.88(1) & 2.96(1) & 1.86(1) & 2.92(1) & 1.82(1) & 2.88(1)  \\
$^8$Li & 2.31(1) & 2.73(1) & 2.19(1) & 2.65(1) & 2.09(1) & 2.45(1) & 2.11(1) & 2.51(1) & 2.07(1) & 2.43(1)  \\
$^8$Be & 2.42(1) & 2.42(1) & 2.48(0) & 2.48(0) & 2.44(1) & 2.44(1) & 2.48(1) & 2.48(1) & 2.39(1) & 2.39(1)  \\
\hline                                                                                                     
$^7$n  &         & 3.08(1) &         & 3.09(1) &         & 2.92(0) &         & 2.85(1) &         & 2.85(1)  \\
$^8$n  &         & 2.98(0) &         & 3.03(1) &         & 2.92(0) &         & 2.88(0) &         & 2.88(1)  \\
\end{tabular}
\label{table:pn-radii}
\end{table}
         
\begin{table}
\caption{Isoscalar and isovector magnetic moments, calculated in impulse
approximation, in nuclear magnetons.}
\begin{tabular}{ccrrrrrr}
              &    T          &  $\avep$    &    UIX~~~  &    IL2~~~  &    IL3~~~   &    IL4~~~   & Expt     \\
\hline                                                                                            
              &               &  \multicolumn{6}{c}{Isoscalar}  \\                                
$^3$He$-^3$H  & $\case{1}{2}$ & 0.408(0)    & 0.405(0)   &   0.403(0) &    0.402(0) &    0.404(0) &   0.426  \\
$^6$Li        &    0          & 0.823(1)    & 0.821(1)   &   0.817(1) &    0.810(1) &    0.819(1) &   0.822  \\
$^7$Be$-^7$Li & $\case{1}{2}$ & 0.904(6)    & 0.90(1)~   &   0.894(1) &    0.895(1) &    0.898(1) &   0.929  \\
$^8$B$-^8$Li  &    1          & 1.31(1)~    & 1.307(2)   &   1.276(1) &    1.287(1) &    1.295(1) &   1.345  \\
\hline                                                                                            
              &               &  \multicolumn{6}{c}{Isovector}  \\                                
$^3$He$-^3$H  & $\case{1}{2}$ & $-$4.354(1) & $-$4.340(1)& $-$4.330(1)& $-$4.316(1) & $-$4.331(1) & $-$5.107 \\
$^7$Be$-^7$Li & $\case{1}{2}$ & $-$3.92(1)~ & $-$3.96(1)~& $-$3.93(1)~& $-$3.84(1)~ & $-$3.96(1)~ & $-$4.654 \\
$^8$B$-^8$Li  &    1          &    0.39(1)~ &    0.40(2)~&    0.369(9)&    0.34(1)~ &    0.377(9) & $-$0.309 \\
\end{tabular}
\label{table:magmom}
\end{table}

\begin{table}
\caption{Quadrupole moments, calculated in impulse approximation, in fm$^2$.}
\begin{tabular}{lccccccc}
       &  $\avep$   &    UIX     &    IL2     &    IL3     &    IL4     & Expt     \\
\hline                                                                               
$^6$Li & $-$0.27(8) & ~$-$0.1(2) & $-$0.32(6) & $-$0.35(6) &   ~0.27(5) & $-$0.083~~  \\
$^7$Li & $-$3.6(1)~ & ~$-$4.5(1) & $-$3.6(1)~ & $-$3.4(1)~ & $-$3.9(1)~ & $-$4.06~~~  \\
$^7$Be & $-$6.4(1)~ & ~$-$7.5(1) & $-$6.1(1)~ & $-$5.4(1)~ & $-$6.6(1)~ &             \\
$^8$Li &  ~~3.5(3)~ &   ~~3.0(1) &   ~3.2(1)~ &  ~~3.2(1)~ &   ~3.4(1)~ &   ~3.19(7)  \\
$^8$B  &  ~~7.1(3)~ &   ~~8.2(1) &   ~6.4(1)~ &  ~~5.6(1)~ &   ~6.6(1)~ &   ~6.8(2)~  \\
\end{tabular}
\label{table:quadrupole}
\end{table}

\begin{table}
\caption{Expectation values of $\dv$ and the net change in the binding energy due to the boost correction
(in MeV) for three of the Illinois Hamiltonians.}

\begin{tabular}{ldddddd}
                        &   \multicolumn{3}{c}{$\dv$}      & \multicolumn{3}{c}{$\dv + (\Vijk^*-\Vijk)$} \\
                        & AV18/IL1 & AV18/IL4 & AV18/IL5   &  AV18/IL1   & AV18/IL4    & AV18/IL5      \\
\hline
$^3$H($\case{1}{2}^+$)  & 0.406(4) & 0.394(4) & 0.395(4)   &    0.000(8) & $-$0.001(8) & $-$0.012(8)   \\
$^3$He($\case{1}{2}^+$) & 0.394(4) & 0.382(4) & 0.386(4)   & $-$0.003(8) &    0.001(8) & $-$0.005(8)   \\
$^4$He(0$^+$)           & 2.13(2)  & 2.09(2)  & 2.08(2)    & $-$0.10(4)  & $-$0.08(4)  & $-$0.10(4)     \\
$^6$He(0$^+$)           & 2.81(9)  & 2.93(8)  & 2.83(8)    &    0.1(1)   &    0.1(1)   &    0.1(1)     \\
$^6$He(2$^+$)           &          &          & 2.78(8)    &             &             &    0.1(1)     \\
$^6$Li(1$^+$)           & 2.81(8)  & 2.96(9)  & 3.04(8)    &    0.2(1)   &    0.2(1)   &    0.2(1)     \\
$^7$Li($\case{3}{2}^-$) & 3.9(1)   & 4.0(1)   & 3.8(1)     &    0.1(2)   &    0.4(2)   &    0.3(2)     \\
$^7$Li($\case{1}{2}^-$) &          &          & 4.1(1)     &             &             &    0.4(2)     \\
$^7$Li($\case{7}{2}^-$) &          &          & 4.1(1)     &             &             &    0.5(2)     \\
$^8$He(0$^+$)           & 4.2(1)   & 4.5(2)   & 4.5(1)     &    0.5(2)   &    0.0(2)   &    0.2(2)     \\
$^8$Li(2$^+$)           & 5.4(1)   & 5.9(2)   & 5.3(2)     &    0.3(2)   &    0.3(2)   &    0.3(2)     \\
$^8$Li(1$^+$)           &          &          & 5.2(2)     &             &             &    0.1(2)     \\
$^8$Li(3$^+$)           &          &          & 5.7(2)     &             &             &    0.2(3)     \\
$^8$Li(4$^+$)           & 5.0(1)   &          & 5.3(2)     &    0.7(2)   &             &    0.6(2)     \\
$^8$Be(0$^+$)           & 5.6(2)   & 5.6(2)   & 5.8(3)     &    0.1(3)   &    0.3(2)   &    0.3(4)     \\
$^8$Be(1$^+$)           &          &          & 5.3(2)     &             &             &    0.3(3)     \\
\hline
$^8$n(0$^+$)            &          & 1.25(4)  & 1.12(5)    &             &    0.61(5)  &    0.57(6)    \\
\end{tabular}
\label{table:dv}
\end{table}

\begin{figure}
\caption{Energies of ground or low-lying excited states of light nuclei
computed with the AV18 and AV18/UIX interactions, compared to experiment.
The light shading shows the Monte Carlo statistical errors.
The dashed lines indicate the thresholds against breakup for each model
or experiment.}
\label{fig:oldH}
\end{figure}

\begin{figure}
\caption{Three-body force Feynman diagrams.  The first, a, is the 
Fujita-Miyazawa, b is two-pion S-wave, c and d are three-pion rings
with one $\Delta$ in intermediate states.}
\label{fig:vijk}
\end{figure}

\begin{figure}
\caption{Energies computed with the AV18/UIX, AV18/IL2, and AV18/IL4
Hamiltonians compared to experiment for narrow states.  The light
shading shows the Monte Carlo statistical errors.
The dashed lines indicate the thresholds against breakup for each model
or experiment.}
\label{fig:energies}
\end{figure}

\end{document}